\documentclass[letterpaper,twocolumn,10pt]{article}
\usepackage{usenix2019_v3}

\usepackage{tikz}

\usepackage{cite}
\usepackage{amsmath,amssymb,amsfonts}
\usepackage{graphicx}
\usepackage{textcomp}
\usepackage{xcolor}

\usepackage{booktabs}
\usepackage{array}

\usepackage{algorithm}
\usepackage{algpseudocode}
\usepackage{xspace}
\usepackage{subcaption}
\usepackage{multirow}
\usepackage{comment}
\newcommand{\contentionratio}{\rho_t}

\usepackage{color}
\usepackage{soul}

\usepackage{tabularx}

\newcommand{\note}[1]{\textcolor{red}{{#1}}}

\newcommand{\circled}[1]{\tikz[baseline=(char.base)]{\node[shape=circle,draw,inner sep=1pt,font=\scriptsize,fill=gray!20] (char) {#1};}
}

\begin{document}
%-------------------------------------------------------------------------------

%don't want date printed
\date{}

%\title{\Large \bf Network Contention State Aware Dynamic Optimizer for Accelerator Collective Communication Libraries}

\title{\Large \bf Network Contention-Aware Dynamic Optimizer for Accelerator Collective Communication}

%for single author (just remove % characters)
% \author{\note{Anonymous Authors}}

\author{
{\rm Rajat Bisht}\\
Tennessee Technological University
\and
{\rm Tevfik Kosar}\\
University at Buffalo, Buffalo
\and
{\rm Anthony Skjellum}\\
Tennessee Technological University
\and
{\rm Gene Cooperman}\\
Northeastern University
\and
{\rm MD S Q Zulkar Nine}\\
Arkansas Tech University
}

\maketitle

%-------------------------------------------------------------------------------
\begin{abstract}
%-------------------------------------------------------------------------------

Collective Communication Libraries (CCL) are critical for large-scale distributed AI/ML training, yet their configurations are determined by static heuristics that do not account for real-time network conditions. 
Our empirical study shows that these heuristics perform near-optimally in contention-free environments, but degrade substantially under background traffic. 
To address this, we present a cooperative multi-agent run-time optimizer for accelerator collective communication. Our approach employs a two-agent architecture: (i) {\em detector} -- a Long Short Term Memory (LSTM) recurring neural network fed classifier head with a sequential probability ratio test that distinguishes sustained contention from transient spikes; 
and (ii) {\em selector} -- a configuration selector built on a Dueling Double DQN (D3QN) with Conservative Q-Learning that adapts configuration parameters in response. 
We implement this dynamic tuner over NCCL, a widely used CCL developed by NVIDIA. 
When evaluated on Chameleon Cloud's isolated bare-metal cluster, this dynamic tuning approach reduces AllReduce latency by 36\% over NCCL and 32\% over a fully converged AutoCCL. 
We further evaluate it across six real AI workloads spanning convolutional networks, transformers, and recommendation architectures. 
Our multi-agent tuner approach delivers up to 65\% speedups on Delta cluster with NVIDIA A100 GPUs and Slingshot-11 fabric internode connectivity.

\end{abstract}

%-------------------------------------------------------------------------------
% The sections 
%-------------------------------------------------------------------------------

\section{Introduction}
%%====================================================================

% WHO WOULD CARE?
The rapid proliferation of AI/ML workloads and LLM-based agentic systems have driven model sizes to scales that fundamentally exceed the memory and compute capacity of an individual accelerator, both during training as well as during inference~\cite{demystifying_nccl, pccl, msccl++,zeus, eff_schedule}. 
To overcome this constraint, researchers and industry practitioners have turned to distributed GPU architectures, employing various training paradigms such as data parallelism (DP)~\cite{dp_horovod,dp_pytorch,dp_zero}, tensor parallelism (TP)~\cite{tp_megatron_lm,tp_megatron_lm_2,tp_2d_method}, pipeline parallelism (PP)~\cite{pp_gpipe,pp_pipedream,pp_chimera}, and hybrid combinations thereof~\cite{hyb_depspeed_megatron,hyb_alpa,hyb_gspmd}. 
These paradigms require periodic synchronization using collective communication operations~\cite{demystifying_nccl,msccl++,ncclX} (e.g., \textit{AllReduce}, \textit{ReduceScatter}, \textit{AllGather}), invoked repeatedly to exchange gradients during training, and intermediate results during inference. 
The resulting communication overhead is substantial, with collective operations accounting for up to 80\% of total execution time in large model training~\cite{comm_vs_comp_scaling_2} and nearly two-thirds of a typical training cycle~\cite{pccl}.

%% Prior works and their limitations
Prior efforts to address the communication bottleneck in distributed GPU training fall into the following three broad categories: 
(a) efficient communication algorithms (e.g., topology-aware~\cite{swing,mscclang}, Pareto-optimal~\cite{sccl}, data compression~\cite{hipress,comm_vs_comp_scaling_2,agarwal2022utility}, quantization~\cite{zero_pp}); 
(b) computation-communication overlap techniques~\note{\cite{dp_horovod,pp_gpipe, pp_pipedream,pp_chimera}}; and 
(c) low-level optimized communication library stacks~\cite{nccl, rccl, msccl, msccl++, ncclX}. 
We identify a fundamental limitation common to all the three categories: an implicit assumption that network conditions are stationary, leading to systematically suboptimal communication decisions under real-world network contention. 
This observation motivates us to explore a novel category of optimization, which
we instantiate on NCCL~\cite{nccl} given its status as the de facto standard in distributed AI training and its growing adoption within the HPC community.

%%====================================================================
% PROBLEM DEFINITION

The stationary network assumption manifests in two distinct limitations. 
First, heuristic-based statically tuned libraries (like NCCL and RCCL) select communication configurations optimized for peak hardware bandwidth and minimum latency~\cite{demystifying_nccl,ncclX}, without any mechanism to detect or respond to changes in available bandwidth and transmission delays due to network contention at runtime. 
As a result, when network contention arises, these libraries continue operating under configurations chosen for ideal conditions and systematically under-perform, unaware of the degraded network state.

Second, run-time tuning approaches for online searching of optimal configuration parameter (such as AutoCCL~\cite{autoccl}) address this static heuristics limitation by measuring runtime bandwidth and latency telemetry of initial few cycles. 
However, this approach carries significant tuning overhead and captures only a \emph{static} snapshot of network conditions at startup time. 
This also leads to collective communication performing extremely poorly during exploration cycles due to bad configurations as well as online tuning overhead.
More critically, the chosen configuration grows increasingly suboptimal as network contention evolves. 
Together, these limitations reveal a fundamental measurement bias: existing solutions optimize for network conditions they observe at startup, rather than the conditions actually encountered at run-time.

%%====================================================================
% PROPOSED SOLUTION
%

We propose a novel network contention aware dynamic tuning optimizer for accelerator collective communication libraries, that explicitly models stochastic network contention rather than assuming stationary conditions.
This dynamic tuning optimizer consists of two cooperative intelligent agents trained offline on latency traces collected from production grade clusters, ensuring the learned models capture clean network behavior free from background traffic interference. 
The 'detector' agent continuously monitors collective communication latency at runtime to infer the current contention state of the network and records current network state in LSTM embeddings. 
The 'selector' agent consumes these network state embeddings and selects a NCCL configuration for the current conditions using learned policy, thus incurring no tuning overhead at runtime. 
To account for the uniqueness of each deployment, whether bare-metal or shared, and across varying network fabrics (e.g., standard Ethernet, RoCE, InfiniBand, Slingshot), the dynamic tuning optimizer supports a one-time, lightweight cluster calibration step that adapts the pre-trained models to cluster-specific characteristics. 
Once deployed, the dynamic tuning optimizer integrates seamlessly into existing training stacks using NCCL's Tuner Plugin API. 
By design, both agents run inference on CPU rather than GPU, imposing no burden on training accelerators.

We evaluate our tuner on two clusters: a) Chameleon Cloud~\cite{cc}, providing isolated bare-metal nodes for controlled microbenchmark and AI workload evaluation; and b) Delta Cluster with A100 GPUs and the Slingshot-11 interconnect for scaling validation.
On Chameleon, the dynamic tuning optimizer achieves an average latency reduction of 36\% over NCCL and 32\% over a fully converged AutoCCL, with substantially lower latency variance than both baselines, yielding more stable and predictable training iterations.
We further evaluate across six real AI training workloads: ResNet-50~\cite{resnet}, BERT~\cite{bert}, NanoGPT~\cite{nanogpt}, DLRM~\cite{dlrm}, GPT-2 Medium~\cite{gpt2}, and ViT-Large~\cite{vit}.
Scaling to over 24 nodes on Delta's Slingshot-11 fabric under real-world network contention, the dynamic tuning optimizer achieves speedups up to 65\%.
A detailed evaluation is presented in Section~\ref{sec:eval}.

%%====================================================================
% CONTRIBUTIONS
\noindent This paper makes the following contributions:
\begin{enumerate}
\vspace{-0.1in}
\item {\em Network-State-Aware Collective Communication.} We introduce a new class of collective communication optimizations that models for stochastic network contention state instead of considering it stationary. The dynamic tuning optimizer realizes this as a deployable system selecting NCCL configurations at runtime with negligible overhead.
\vspace{-0.1in}
\item {\em Decoupled Coordinator-Plugin Architecture.} We propose a two-tier MPI-based skeleton-stub coordinator hierarchy that offloads configuration selection entirely from the NCCL critical path to a separate process, scaling with the number of nodes participating in NCCL workload rather than the number of NCCL ranks.
\vspace{-0.1in}
\item {\em Sustained Contention Detection.} We design contention detector, which observes collective communication latency and employs a novel approach of using long short-term memory (LSTM)~\cite{lstm} recurring neural network to detect anomaly in network state. This network anomaly is then classified using a classifier head combined with a sequential probability ratio test (SPRT~\cite{sprt}) to strongly distinguish sustained contention, transient spikes, and contention-free states with low detection latency.
\vspace{-0.1in}
\item {\em Congestion-Aware Configuration Selection.}
We design configuration selector agent, a conservative dueling double deep Q-learning agent that operates with network state awareness using LSTM raw embeddings and inferred contention states to identify new NCCL tuning configurations.
\vspace{-0.1in}
\item {\em Constant-Time Critical Path.}
We cache the configuration decisions from configuration selector agent in a KV store to reduce duplication of efforts and amortize inference call benefits.
\vspace{-0.1in}
\item {\em CPU-Only Agent Inference with Negligible Overhead.} Both agents run inference exclusively on CPU, imposing no additional burden on training accelerators, with a combined inference latency of 3-4 ms.
\end{enumerate}

\section{Background and Motivation}
\label{sec:bkgrd_motiv}

%%==========================================================
\subsection{Collective Communication Primitives}
\label{sec:coll_comm}

%% Collecitve Primitives disection
Collective communication operations are synchronized primitives requiring all participating processes to coordinate jointly to complete a well-defined communication task~\cite{survey_coll_optim}.
These operations fall into two broad categories:
(a) communication-only operations such as \textit{Broadcast} and \textit{AllGather}, which redistribute data across processes without modification, and
(b) reduction-communication operations such as \textit{AllReduce} and \textit{ReduceScatter}, which combine reduction computations with data exchange.
Collective operations are realized through two primary algorithmic families:
(a) ring-based algorithms offering bandwidth-optimal performance; and
(b) tree-based algorithms for latency sensitive implementations.
These collective algorithms operate on Simple and Low-Latency protocols across shared memory for intra-node and RDMA for inter-node transport.
Together, these primitives form the synchronization backbone of distributed computing, ensuring all participating processes reach a consistent shared state before proceeding.

\begin{figure}[t]
    \begin{subfigure}{0.49\linewidth}
        \includegraphics[width=\linewidth, height=45mm]{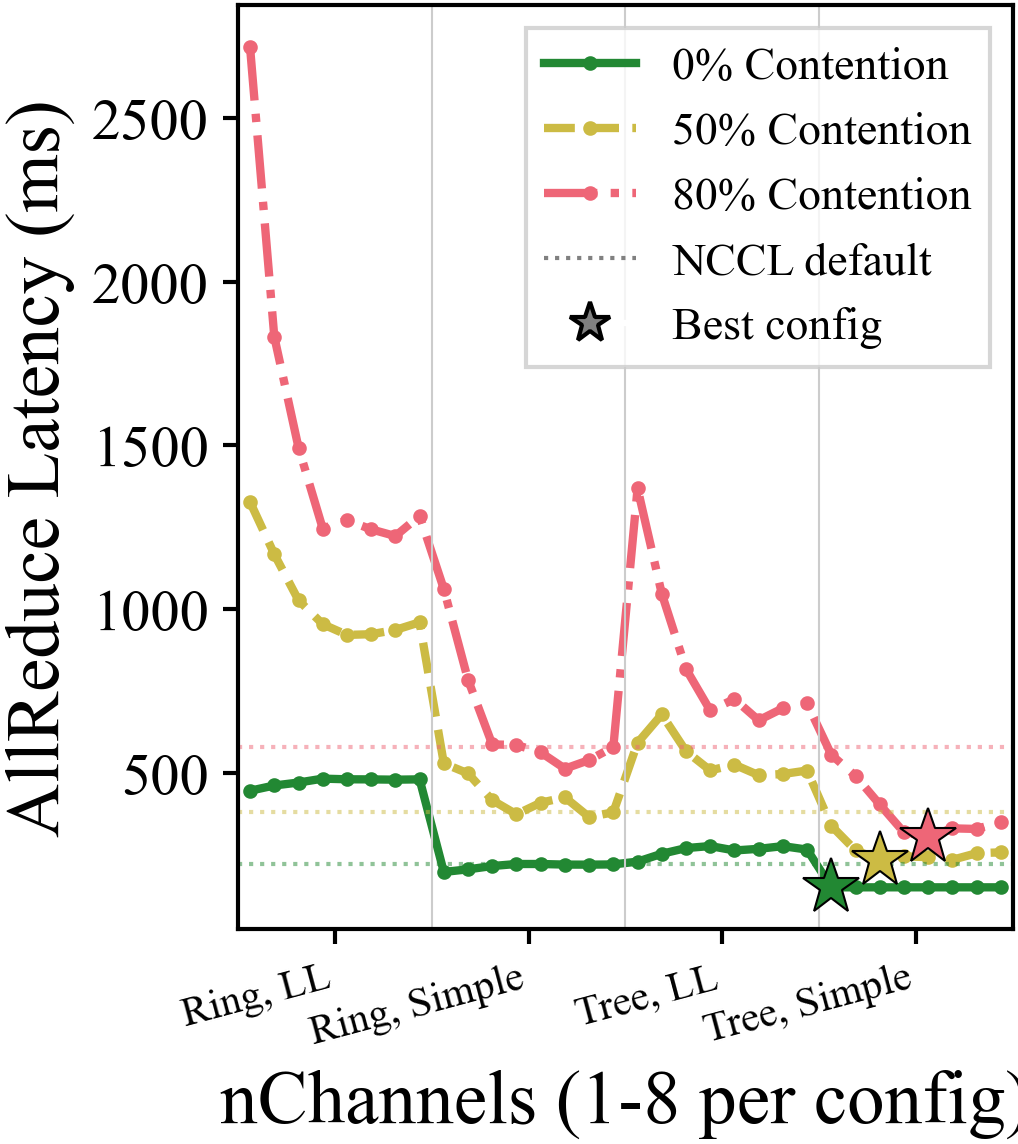}
        \caption{256MB}
    \end{subfigure}
    \begin{subfigure}{0.49\linewidth}
        \includegraphics[width=\linewidth,height=45mm]{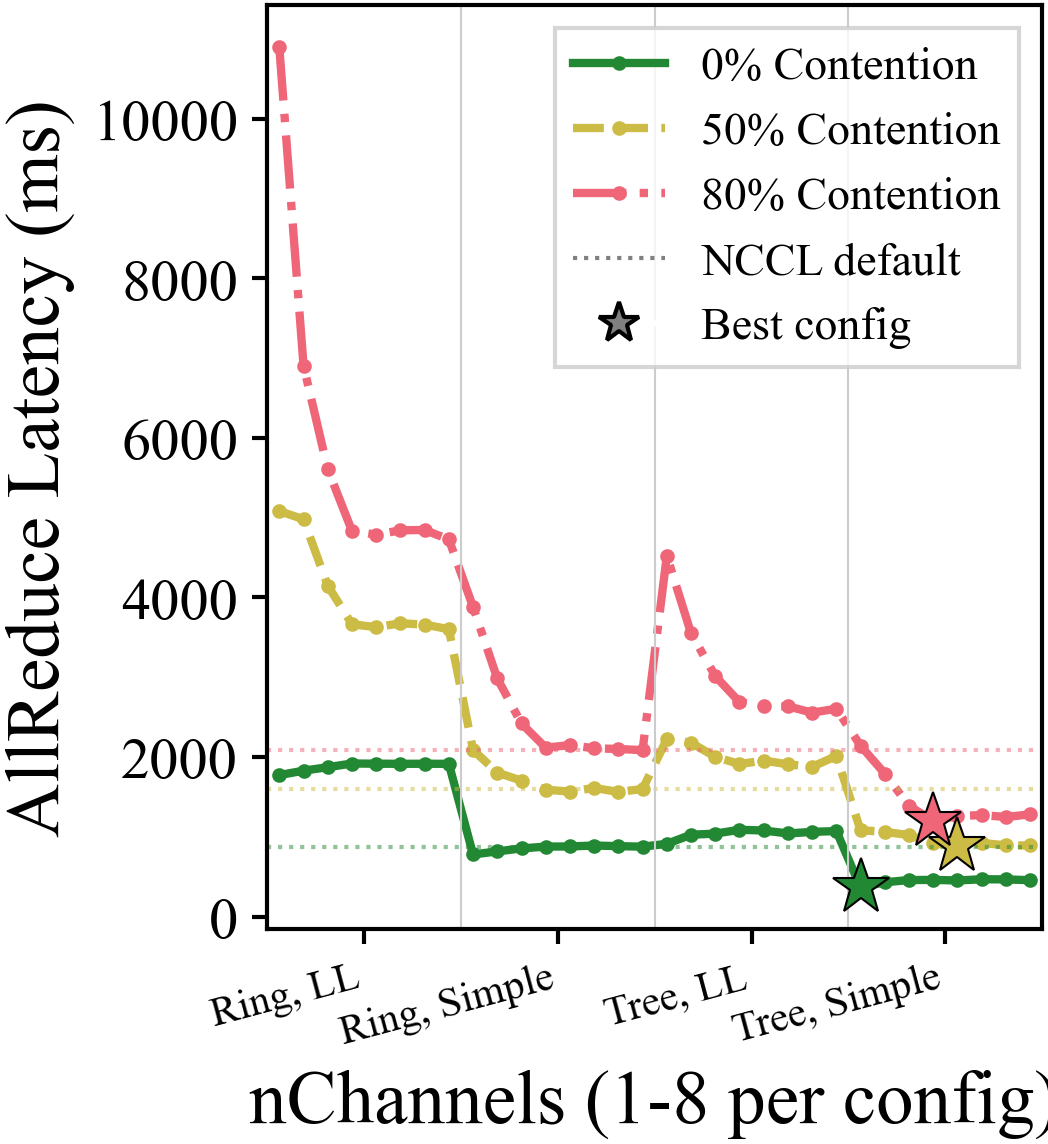}
        \caption{1GB}
    \end{subfigure}
    \caption{NCCL configuration sweep on 25 Gbps duplex network under various levels of contention.}
    \vspace{-2mm}
    \label{fig:nccl_config_sweep}
\end{figure}

%% Collective Communication Libraries (NCCL, RCCL, MSCCL, CUDA-aware MPI)
To efficiently execute these primitives in modern GPU-accelerated clusters, specialized communication libraries and interfaces have been developed.
While AMD’s ROCm Communication Collectives Library (RCCL)~\cite{rccl} provides optimized primitives for AMD accelerators, Microsoft’s MSCCL~\cite{msccl} introduces programmable collective topologies, and CUDA-aware MPI implementations~\cite{cuda_aware_mpi} facilitate direct GPU-to-GPU transfers in traditional HPC environments, the NVIDIA Collective Communications Library (NCCL)~\cite{nccl} has emerged as the de facto industry standard backend~\cite{gin}.
Because mainstream deep learning frameworks such as PyTorch~\cite{dp_pytorch} rely on NCCL as default backend for distributed training and inference across modern GPU clusters, low-level collective optimizations built on NCCL offer the most direct and impactful performance gains for large-scale AI workloads.
Consequently, this work adopts NCCL as the foundational target for our optimizations.

%% use of Collectives in AI training and inference
Distributed AI training and inference rely fundamentally on collective communication to synchronize state across accelerators.
Data, tensor, and pipeline parallelism each invoke \textit{AllReduce}, \textit{ReduceScatter}, and \textit{AllGather} repeatedly, either to aggregate gradients, synchronize activations, or pass results across pipeline stages~\cite{msccl++, ncclX}.
Modern LLM frameworks further decompose \textit{AllReduce} into \textit{ReduceScatter} and \textit{AllGather} to enable memory efficiency and computation-communication overlap.
%
%% Collecitve communciaiton is bottle neck
The volume and frequency of these operations make collective communication the dominant bottleneck at scale.
During LLM inference, communication accounts for 10--40\% of end-to-end execution time, with GPT-3~\cite{gpt3} spending approximately 30\% in \textit{AllReduce} alone~\cite{msccl++}.
During LLM training, LLaMA-2-7B spends up to 80\% of execution time in communication~\cite{comm_vs_comp_scaling_2}, and scaling LLaMA-2-13B~\cite{llama2} from 8 to 1024 GPUs collapses Model FLOPs Utilization (MFU) from 47\% to 4\%~\cite{llama2_overhead} due to delays in communication.
Across models like MicroLlama, OPT, GPT-2, and DistilGPT-2, communication constitutes 23\%--64\% of training time~\cite{pccl}, with collective operations accounting for nearly two-thirds of a typical training iteration~\cite{pccl}.

%%==========================================================
\subsection{NCCL: Internals and Limitations}
\label{sec:nccl_limits}

% -------------------------------------------------------------------------------------------------------------------

% -------------------------------------------------------------------------------------------------------------------

%% NCCL configurations
State-of-the-art collective communication libraries such as NCCL and its various extensions~\cite{ncclX, gin, msccl++} are oblivious to real-time network contention.  
NCCL bridges the CPU processes with the GPU kernels, selecting configurations across three dimensions:
(a) algorithm (Ring, Tree, and hardware-dependent variants requiring NVIDIA SHARP or NVSwitch);
(b) protocol (LL, Simple, and NVLink-dependent LL128); and
(c) number of channels, controlling link parallelism~\cite{demystifying_nccl}.
NCCL runtime logs show that configuration is determined using heuristics with data collected at initialization by probing network interfaces with small messages (8B--4KB) and inferring network bandwidth from the static topology.
For added flexibility, NCCL also allows a Tuner Plugin API enabling users to inject configurations at runtime.

% -------------------------------------------------------------------------------------------------------------------
\begin{figure}[t]
    \begin{subfigure}{0.49\linewidth}
        \includegraphics[width=\linewidth, height=45mm]{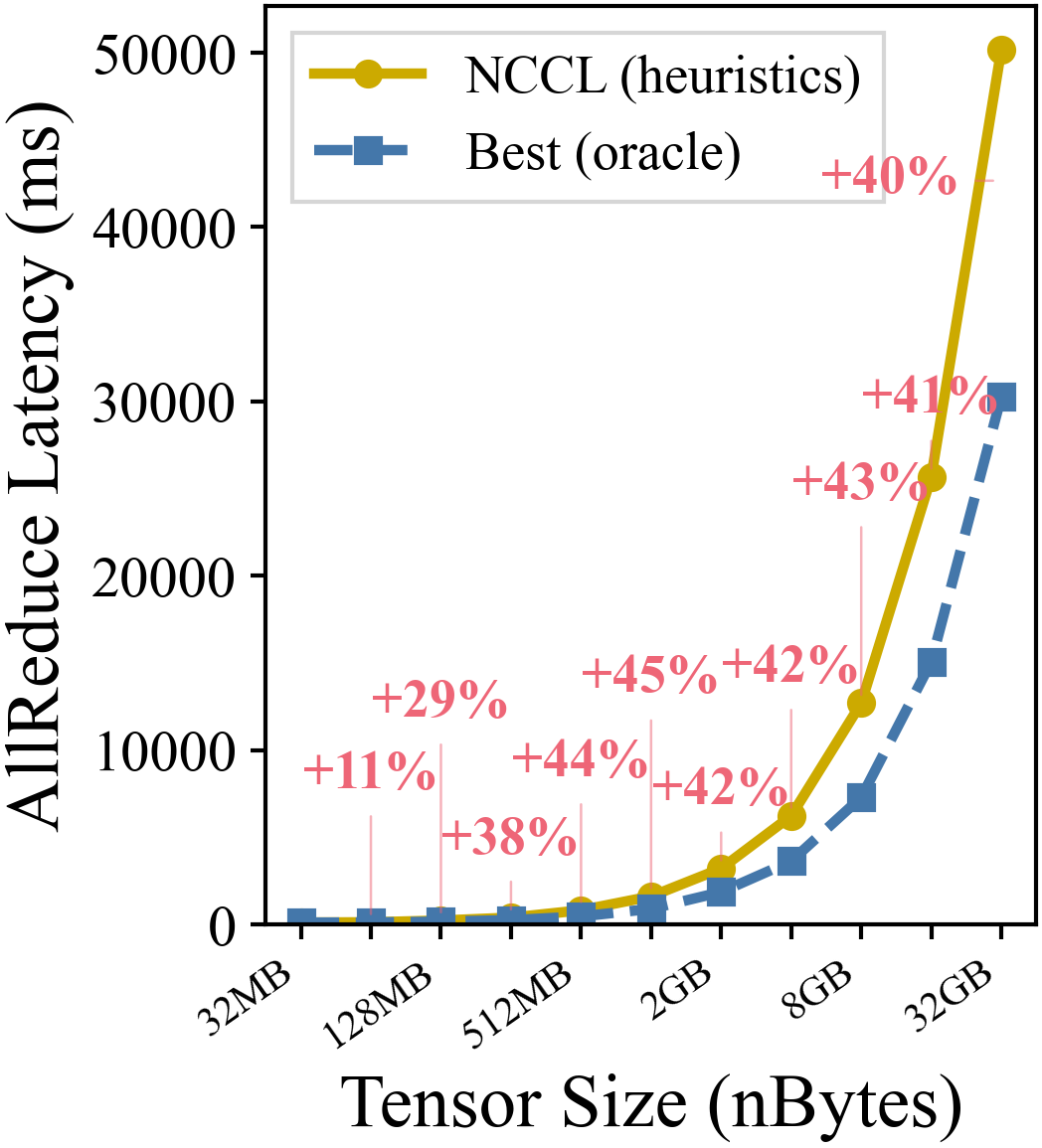}
        \caption{50\% Network Contention}
    \end{subfigure}
    \begin{subfigure}{0.49\linewidth}
        \includegraphics[width=\linewidth,height=45mm]{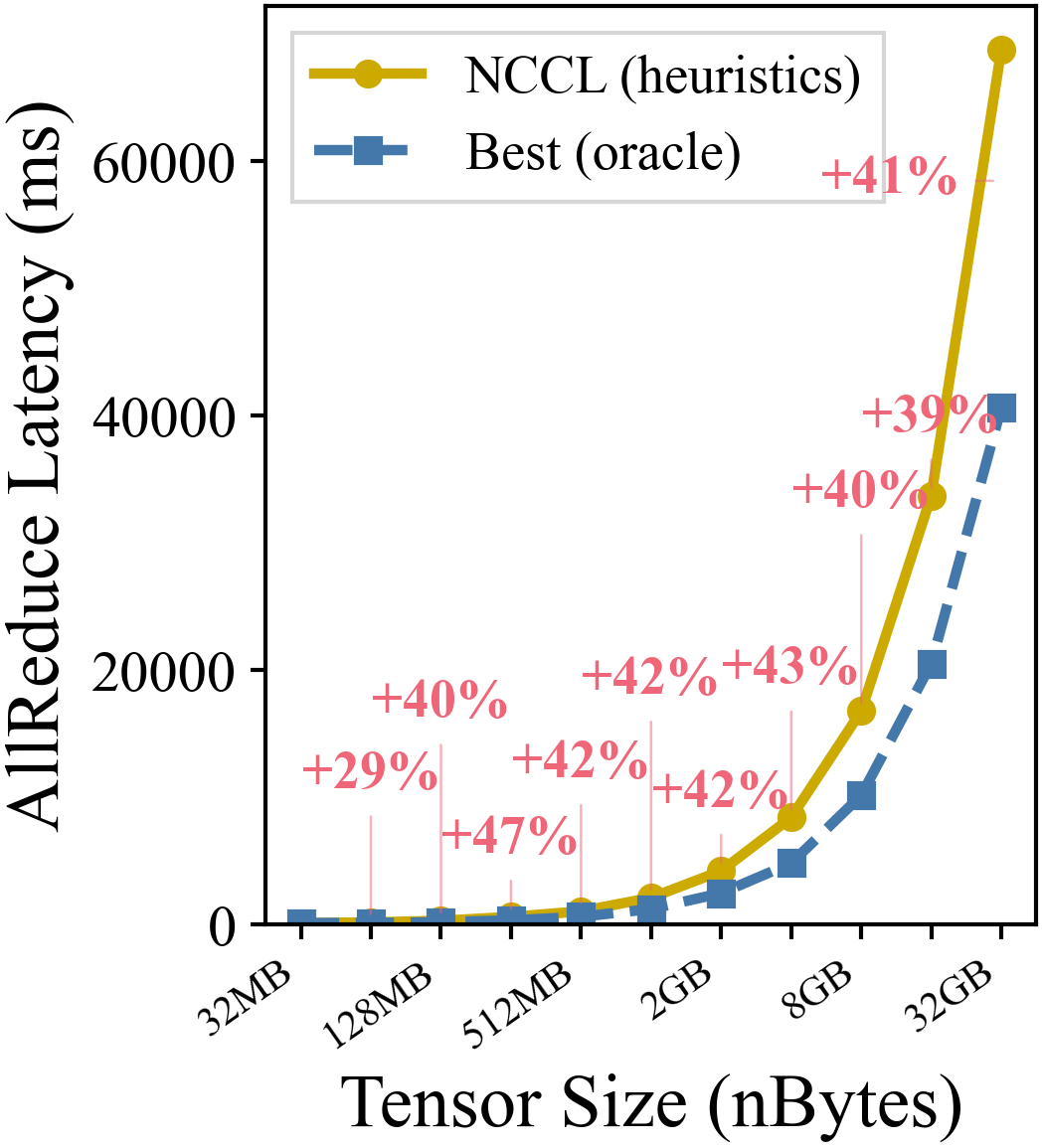}
        \caption{80\% Network Contention}
    \end{subfigure}
    \caption{Comparing NCCL-default configuration with empirically optimal solution under network contentions.}
    \vspace{-2mm}
    \label{fig:nccl_speedup}
\end{figure}

%% NCCL limitations
NCCL's configuration selection rests on two narrow foundations: 
small-message probes that never induce queue buildup or link saturation, and static topology information that reflects nominal state rather than effective bandwidth under contention. 
To assess the optimality of NCCL's heuristic choices, we conduct an exhaustive sweep over its configuration space under varying levels of injected network contention. 
The results of this sweep are summarized in Figures~\ref{fig:nccl_config_sweep} and~\ref{fig:nccl_speedup}.
While NCCL's heuristics remain near-optimal under contention-free conditions, alternative configurations yield 3\%--22\% speedup at 50\% contention and 3\%--52\% at 80\% contention, spanning message sizes from 32\,MB to 32\,GB on a 25\,Gbps duplex link. 
These findings establish that NCCL's heuristics-based configuration choices are systematically suboptimal in the presence of network contention, motivating a runtime-aware approach to configuration selection.

%%==========================================================
\subsection{Runtime Tuning and Its Limitations}
\label{sec:autoccl_limits}

\begin{figure}[b]
    \centering
    \includegraphics[width=1.0\linewidth]{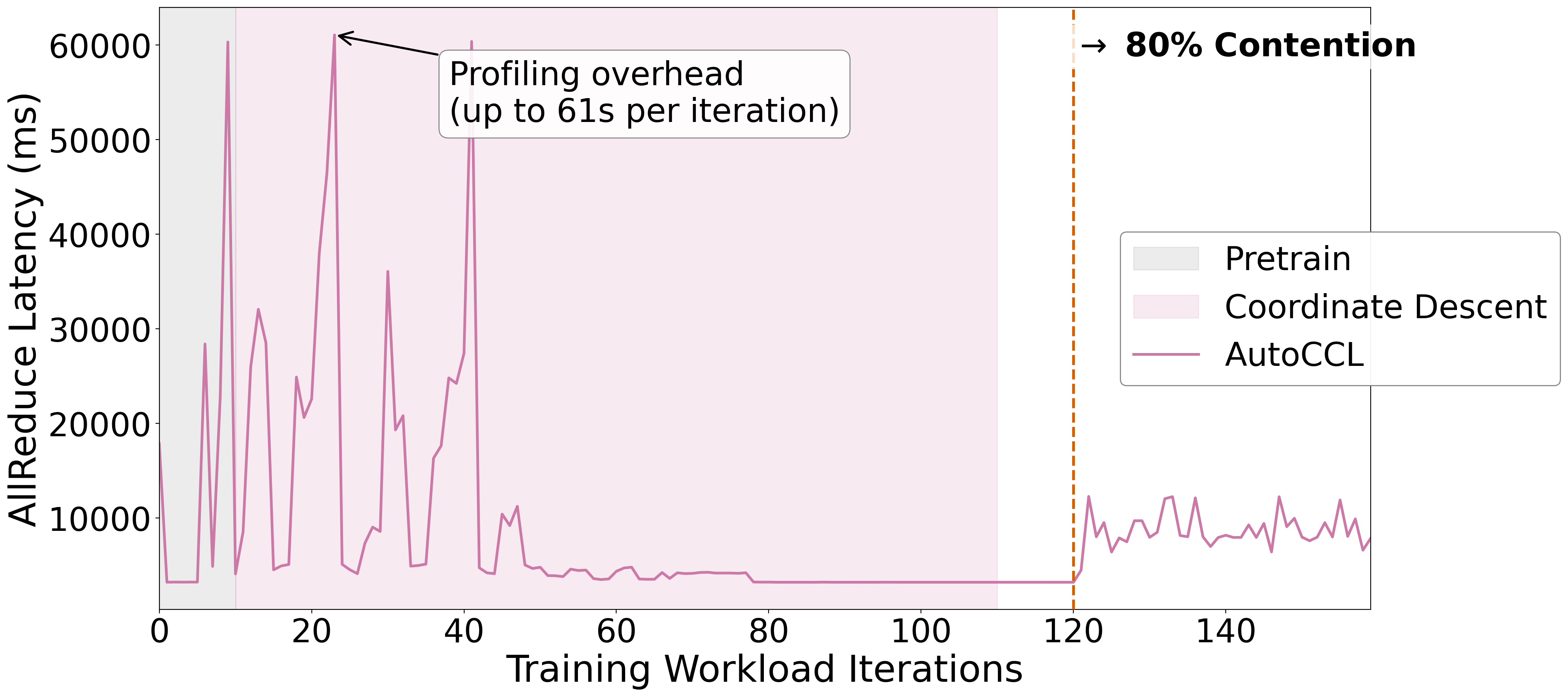}
    \caption{AutoCCL run-time profiling overhead.}
    \vspace{-2mm}
    \label{fig:autoccl_overhead}
\end{figure}

%% AutoCCL
Run time configuration tuning approaches, like the ones implemented by AutoCCL~\cite{autoccl}, challenges the prevailing view that NCCL is inherently well-tuned by showing that substantial performance headroom remains accessible with runtime configuration tuning.
This observation provides the foundational insight that motivates our work. 
By modifying the open-source NCCL library, AutoCCL introduces support for runtime measurements of bandwidth usage and latency of collective communication calls. 
AutoCCL then runs online search by exploring the configuration space using coordinate descent, while incurring a runtime tuning overhead across a predefined number of tuning calls.
As shown in Figure~\ref{fig:autoccl_overhead}, this online search imposes substantial overhead, with per-iteration latency spiking to as much as 61 seconds during the tuning phase.

% -------------------------------------------------------------------------------------------------------------------
% -------------------------------------------------------------------------------------------------------------------
 
Beyond this direct cost, we further examine the sensitivity of online tuning to two qualitatively distinct contention regimes: short-lived transient spikes and sustained contention.
When transient contention occurs during the tuning phase, the optimization trajectory is perturbed, causing convergence to a configuration tuned for a network snapshot that no longer reflects current conditions.
More fundamentally, existing runtime tuners commit to a single configuration upon convergence and do not re-optimize thereafter.
Therefore these online tuning approaches also inherit the same measurement bias as NCCL: 
the selected configuration reflects conditions observed during tuning rather than those encountered over the remainder of execution.

%%==========================================================
\subsection{Network-State-Aware Optimization}
\label{sec:dccl_motiv}

Existing state-of-the-art models treat the network as static, optimizing for conditions observed once and assumed to persist.
In practice, contention on shared multi-user clusters is stochastic and time-varying, driven by competing workloads, background traffic, and transient link saturation.
Any solution that commits to a fixed configuration without tracking network state will inevitably suffer from a measurement bias.

Addressing this gap requires a solution that continuously observes network state, distinguishes sustained contention from transient noise, and selects configurations adaptive to current conditions.
The structural properties of this problem motivate a learning-based optimization framework, which we characterize formally in Section~\ref{sec:proposed_solution}.

\section{Network Contention Aware Dynamic Tuning}
\label{sec:proposed_solution}

%%==============================================

Network-state-aware configuration selection for collective communication is governed by three properties: (a) \textit{stochasticity}, where contention is driven by external workloads that are neither predictable nor controllable; (b) \textit{temporal dependency}, where contention anomalies are better detected with historical context than from instantaneous observations alone; and (c) \textit{sequentiality}, where each configuration decision influences the performance of subsequent collective calls.
Therefore, heuristic- and rule-based approaches cannot adapt to conditions outside of their design envelope.

Reinforcement learning (RL) offers a more principled fit, as an RL agent learns policy from the environment (e.g., network condition) guided by reward accumulation over time, naturally accommodating the sequential and stochastic nature of the problem.
Configuration selection maps directly onto a Markov Decision Process (MDP): the agent observes network state, selects a configuration, receives a reward reflecting collective performance, and transitions to a new state.
However, a single monolithic agent faces a structural tension: a contention detection model requires a recurrent architecture for accumulating evidence across multiple iterations, whereas a configuration selection model requires a policy optimization based decision making.
Therefore, the monolithic model must run inference on every collective call and generate configurations regardless of whether the network state has changed or not, imposing unnecessary overhead on the critical path.

To solve this, we decompose the optimizer into a two-agent cooperative model.
A contention detector agent observes collective call latency over time and classifies the network state into three categories: SUSTAINED-CONTENTION, TRANSIENT-SPIKE, or STEADY-STATE.
A configuration selector agent is then triggered exclusively during SUSTAINED-CONTENTION, using raw embeddings from the detector to produce a network-state-aware configuration.
This decomposition isolates network dynamic detection from configuration optimization, ensuring that inference cost for configuration selection is only incurred when an actionable state change is confirmed.

%%==============================================
\subsection{Temporal Contention Detection }
\label{sec:agent1_temporal}

%% --- temporal dependency: what and why
At each collective communication operation, detector agent addresses a single well-defined question: {\em upon detecting a performance perturbation, is detected anomaly a safely ignorable transient spike, or a reconfiguration warranting persistent state?} To quantify the magnitude of degradation, one of the input to detector agent is a computed contention ratio:
\vspace{-1mm}
\begin{equation}
    \contentionratio = L_{\text{obs}} / L_{\text{baseline}}
    \label{eq:contention_ratio}
    \vspace{-1mm}
\end{equation}

\noindent where $L_{\text{baseline}}$ is the median latency observed during the first few collective operations under zero contention for the participating ranks and nodes (${nRanks}$, ${nNodes}$) executing a particular collective type (${collType}$) for message size (${nBytes}$). Values of $\contentionratio$ near 1.0 indicate zero network contention; values above 1.0 reflect degradation proportional to contention severity.
Because a single $\contentionratio$ value cannot distinguish a transient spike from sustained contention without temporal context and rendering the problem partially observable, a temporal dependency must be incorporated into the state representation.

%% ---- Why LSTM
This temporal dependency motivates the use of an Long Short-Term Memory (LSTM)~\cite{lstm} recurring neural network, which maintains two recurrent states across iterations: a hidden state $h_t$ encoding a short-term summary of recent observations, and a cell state $c_t$ carrying longer-range context through gated updates. Together, $(h_t, c_t)$ accumulate evidence over successive collective calls, enabling the model to distinguish sustained shifts from transient fluctuations.
Although we initially evaluated a recurrent Asynchronous Advantage Actor-Critic (A3C)~\cite{a3c} agent, it suffered from premature action extinction, collapsing spike detection accuracy below 10\%. Given the high variability of shared production clusters, this failure mode was unacceptable, prompting us to adopt a supervised LSTM formulation instead.

%% Why SPRT head
Rather than applying a fixed threshold on contention detector agent's output, we employ the Sequential Probability Ratio Test (SPRT)~\cite{sprt}, which accumulates likelihood ratios sequentially and commits only when sufficient statistical confidence is reached.
With calibrated upper and lower thresholds $\tau_U = 0.95$ and $\tau_L = 0.05$ respectively, the decision rule is:
%\vspace{-2mm}
\begin{equation}
    a_t = \begin{cases}
            \mathrm{SUSTAINED\text{-}CONTENTION} & p_t > \tau_U \\
            \mathrm{TRANSIENT\text{-}SPIKE}     & p_t < \tau_L \\
            \mathrm{STEADY\text{-}STATE}      & \tau_L \leq p_t \leq \tau_U
        \end{cases}
\label{eq:sprt}
%\vspace{-2mm}
\end{equation}

\noindent where $p_t = P(\text{SUSTAINED-CONTENTION} \mid \mathbf{u}_{1:t})$ is the posterior probability of sustained contention given all observations up to iteration $t$.
The SPRT boundaries jointly govern the false-positive rate corresponding to unnecessary reconfigurations, and the false-negative rate corresponding to missed sustained contention detection. 

%% Training results
Network contention detector agent is trained with supervised binary cross-entropy against oracle labels from offline traces and requires no reward signal or online interaction with the environment. Following offline training, the detector achieves 97.5\% accuracy for SUSTAINED-CONTENTION detection, and 84.9\% accuracy for TRANSIENT-SPIKE isolation.

%% agent-1 USP: contention vector for rich netwrok state downstream 
Along with the SUSTAINED-CONTENTION signal, the detector agent also generates a continuous contention vector capturing direction, magnitude, and confidence of the degraded network state. Rather than passing a simple binary alert, this rich representation is forwarded to the configuration selector agent, providing the precise network state context needed to drive downstream reconfiguration.

%%==============================================
\subsection{Contention-Aware Configuration Selection}
\label{sec:agent2_mdp}

%% Configuration action space table for DynamICCL
\begin{comment}
\begin{table}[t]
\centering
\caption{Configuration action space $\mathbb{A}$ for \agenttwo{}. The 96 discrete actions are the Cartesian product of the four dimensions below.}
\label{tab:action_space}
\small
\renewcommand{\arraystretch}{1.2}
\begin{tabular}{@{}llr@{}}
\toprule
\textbf{Symbol} & \textbf{Dimension} & \textbf{Values} \\
\midrule
$\mathrm{A}$    & Algorithm  & Ring, Tree \\
$\mathrm{P}$    & Protocol   & LL, Simple \\
$\mathrm{nCh}$  & Channels   & $1, 2, \dots, 8$ \\
$\mathrm{nTh}$  & Threads    & $128, 256, 512$ \\
\midrule
$|\mathbb{A}|$  & Total actions & $2 \times 2 \times 8 \times 3 = 96$ \\
\bottomrule
\end{tabular}
\end{table}
\end{comment}

\begin{table}[t]
\centering
\caption{Configuration action space $\mathcal{A}$ for Agent-2-OCS. The 96 discrete actions are the Cartesian product of the four dimensions below.}
\label{tab:action_space}
\small
\begin{tabularx}{\linewidth}{l l >{\raggedleft\arraybackslash}X}
\toprule
\textbf{Symbol} & \textbf{Dimension} & \textbf{Values} \\
\midrule
$A$             & Algorithm      & Ring, Tree \\
$P$             & Protocol       & LL, Simple \\
$nCh$           & Channels       & $1, 2, \ldots, 8$ \\
$nTh$           & Threads        & $128, 256, 512$ \\
\midrule
$|\mathcal{A}|$  & Total actions  & $2 \times 2 \times 8 \times 3 = 96$ \\
\bottomrule
\end{tabularx}
\vspace{-4mm}
\end{table}

% --- MDP formualtion
Given the current detected contention state, the configuration selector agent dynamically selects an appropriate NCCL configuration to mitigate communication overhead.
We formulate this as a Markov Decision Process (MDP) $(\mathcal{S}, \mathcal{A}, \mathcal{P}, \mathcal{R})$:

\vspace{-4mm}
\paragraph{ (a) State $s \in \mathcal{S}$:} a 153-dimensional input vector combining contention detector LSTM agent's hidden state $\mathbf{h}_t$ embeddings of accumulated temporal contention history, with instantaneous system telemetry using contention vector $\mathbf{c}_t$, collective parameters, encoded presently selected configuration and normalized GPU statistics (like memory, compute and power usage).

\vspace{-4mm}
\paragraph{ (b) Action $a \in \mathcal{A}$:} A configuration tuple of \{$A$, $P$, $nCh$, $nTh$\} drawn from the 96-element action space defined in Table~\ref{tab:action_space}.

\vspace{-4mm}
\paragraph{ (c) Transition Function $\mathcal{P}$:} Governed jointly by the stochastic evolution of background network contention and the state modifications induced by action $a_t$.

\vspace{-4mm}
\paragraph{ (d) Reward $r \in \mathcal{R}$:} The normalized latency improvement relative to the best zero-contention baseline for the given collective parameters (Equation~\ref{eq:reward})

%% --- Reward fucntion formulation
At each decision step $t$, the reward $r_t$ received by configuration selector agent is defined as:

\begin{equation}
r_t = \frac{L_{\text{zero-best}} - L_{\text{obs}}}{L_{\text{zero-best}}}
\label{eq:reward}
\end{equation}

\noindent where $L_{\text{obs}}$ is the observed collective latency and $L_{\text{zero-best}}$ is the baseline optimal latency under zero contention for the given collective parameters. Because $L_{\text{obs}} \ge L_{\text{zero-best}}$, it makes the reward $r_t \leq 0$, where $r_t = 0$ represents hardware-limit performance. This non-positive formulation grounds the agent's objective in physically achievable baseline performance rather than an arbitrary target.

%% --- D3QN
We adopt a Dueling Double DQN~\cite{ddql_1, ddql_2} architecture, which decouples value and advantage estimation while using a target network to mitigate Q-value overestimation:

\begin{equation}
    y_t = r_t + \gamma (1 - d_t)\, Q_{\theta^-}\!\bigl(s_{t+1}, \arg\max_{a'} Q_\theta(s_{t+1}, a')\bigr)
\label{eq:ddqn}
\end{equation}

\noindent where $\theta$ and $\theta^-$ are the online and target network parameters respectively (with $\theta^-$ slowly updated using soft target updates, $\tau = 0.005$), $\gamma$ is the discount factor, and $d_t$ is a terminal state indicator. To keep inference lightweight, configuration selector agent inherits temporal context directly from contention detector agent's LSTM hidden state $\mathbf{h}_t$ without maintaining its own recurrence and executes in a single feedforward pass.

%% --- CQL
Because training is performed offline on isolated bare-metal cluster traces, the dataset contains an inherent bias: \texttt{TRANSIENT-SPIKE} events are underrepresented as compared to \texttt{SUSTAINED-CONTENTION} and \texttt{STEADY-STATE} conditions. Therefore, without regularization, the agent would overestimate Q-values for the underrepresented state-action pairs. We address this by augmenting the Bellman loss with Conservative Q-Learning (CQL) regularization:

\begin{equation}
\mathcal{L}(\theta) = \mathcal{L}_{\text{Bell}}(\theta) + \alpha_{\text{CQL}}\, \mathcal{L}_{\text{CQL}}(\theta)
\label{eq:cql}
\end{equation}

\noindent where $\mathcal{L}_{\text{CQL}}$ penalizes Q-values for out-of-distribution actions, ensuring the agent defaults to conservative, safe configurations when encountering unfamiliar network states.

%%==============================================
\subsection{Decoupled Hierarchical Coordinator}
\label{sec:cord_plugin_why}

Each NCCL rank loads a tuner plugin that interposes on every collective operation call to inject configuration parameters. Embedding the optimizer directly within this plugin would place agent inference on the critical path of every collective call, adding optimization overhead directly to communication latency. Furthermore, running optimization per rank introduces redundant computation, as co-located ranks evaluate identical network states and reach identical decisions. We resolve both issues by relocating the optimizer to a dedicated coordinator process. The per-rank plugin is thus reduced to a lightweight relay that forwards collective parameters to the coordinator and receives configuration decisions, successfully decoupling optimization overhead from total rank count.

\begin{figure}[t]
\centering
\resizebox{\columnwidth}{!}{
\begin{tikzpicture}[font=\large]

  % Timeline arrow
  \draw[->, thick] (0,0) -- (12,0) node[right] {Time};

  % AllReduce(N)
  \draw[thick] (0.5, 0.15) -- (0.5, -0.15);
  \node[above] at (0.5, 0.15) {\textbf{$CollType$($N$)}};
  \draw[thick, ->] (0.5, -0.4) -- (0.5, -0.9)
    node[below, align=center] {coord replies\\with cached\\config $C$};

  % fwd+bwd pass phase
  \draw[thick, fill=blue!10] (1.5, -0.15) rectangle (10, 0.25);
  \node at (5.5, 0) {fwd pass $\rightarrow$ bwd pass \quad (100ms -- seconds)};
  \draw[thick, ->] (4.0, -0.4) -- (4.0, -0.9)
    node[below, align=center] {coord reads metrics\\triggers agents\\Agent-1 + Agent-2\\run here};

  % AllReduce(N+1)
  \draw[thick] (10.5, 0.15) -- (10.5, -0.15);
  \node[above] at (10.5, 0.15) {\textbf{$CollType$($N$+1)}};
  \draw[thick, ->] (10.5, -0.4) -- (10.5, -0.9)
    node[below, align=center] {coord replies\\with \textbf{updated}\\config $C'$};

\end{tikzpicture}
}
\caption{Agent inference runs during the forward and backward pass compute phase,
hiding optimization latency behind work the application performs regardless.
Config $C'$ is available in the KV cache before $CollType$($N$+1) arrives.}
\vspace{-2mm}
\label{fig:latency_hiding}
\end{figure}
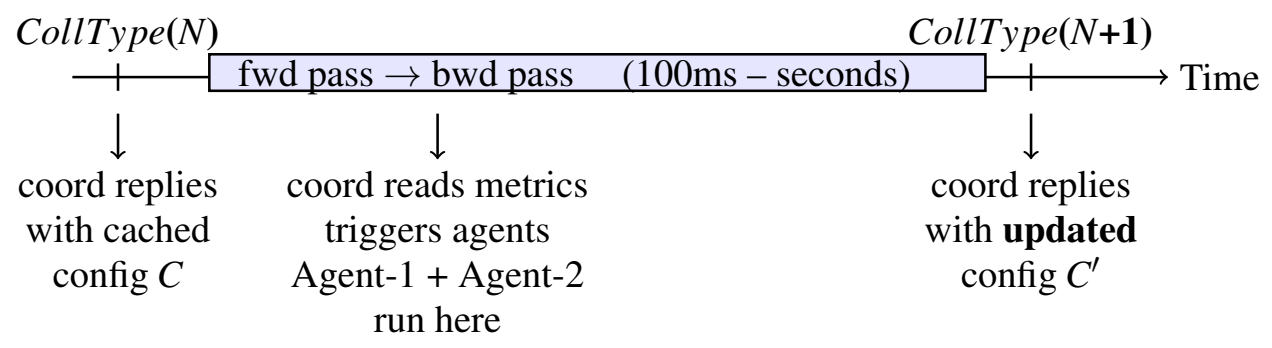

Decoupling the coordinator also enables completely hiding optimization latency through compute-overlap execution. When a collective call occurs, the coordinator replies immediately with a constant-time lookup to a local configuration cache, returning the currently active configuration $C$. It then triggers agent inference asynchronously in the background. As illustrated in Figure~\ref{fig:latency_hiding}, the agents execute during the AI application's forward and backward computation passes which take anywhere from tens of milliseconds to seconds. This allows contention detector and configuration selector agents to complete and update the local cache with new configuration $C'$,  well before the next collective operation ($CollType(N+1)$) arrives. The resulting one-step configuration lag is absorbed entirely within the application's compute time, removing optimization overhead from the critical path.
compute time.

% -----------------------------------------------------------------------
\begin{figure*}[t]
    \centering
    \includegraphics[width=2.0\columnwidth, height=\columnwidth]{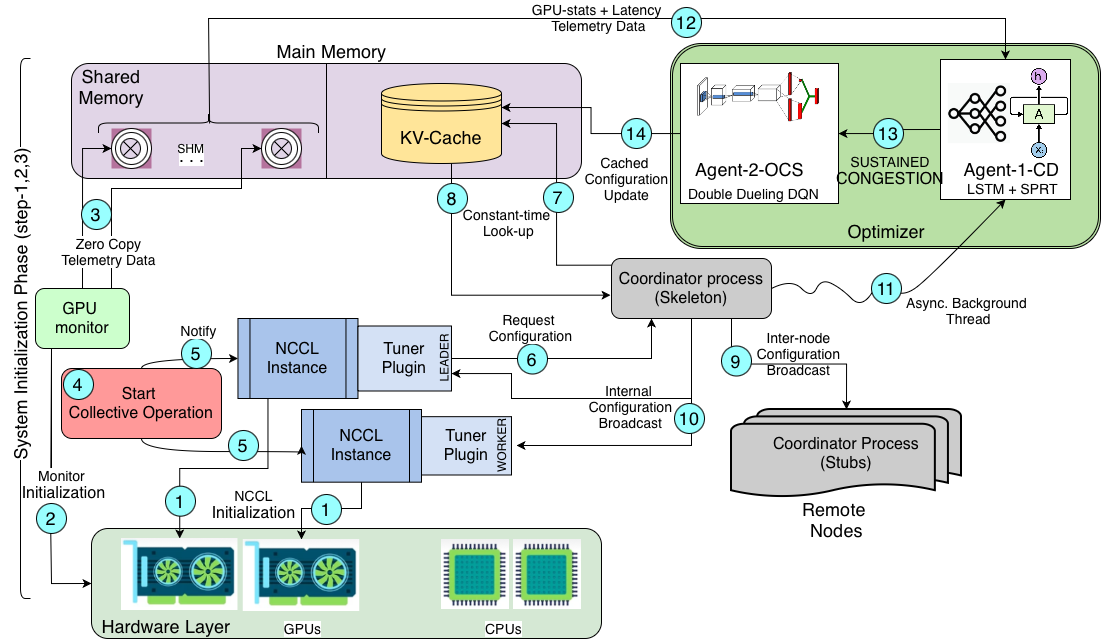}
    \caption{Overview of DynamICCL Architecture.}
    \vspace{-2mm}
    \label{fig:dccl_detailed}
\end{figure*}
% ----------------------------------------------------------
\section{Design and Implementation}
% ---------------------------------------------------------------------

\begin{figure}[b]
    \centering
    \includegraphics[width=1.0\columnwidth]{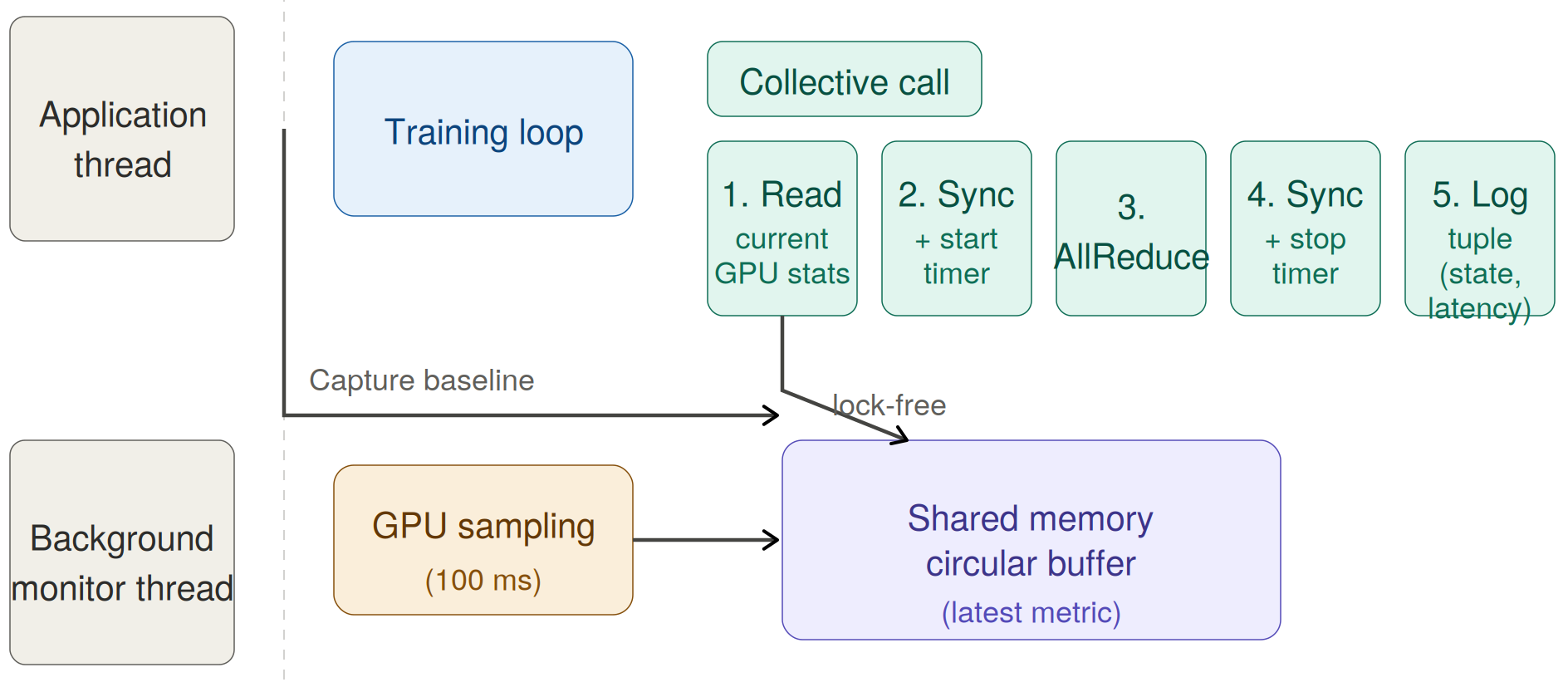}
    \caption{GPU Telemetry and Shared Memory Pipeline}
    \label{fig:python_lib}
\end{figure}

\begin{figure}[t]
    \centering
    \includegraphics[width=1.0\columnwidth]{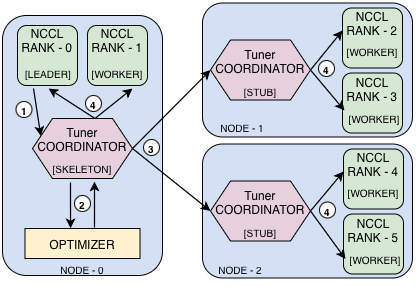}
    \caption{Two-tier coordinator control flow: (1)~configuration request from \texttt{LEADER} rank, (2)~optimizer lookup, (3)~inter-node broadcast from \texttt{SKELETON} to \texttt{STUB} coordinators, and (4)~intra-node dispatch to per-rank \texttt{WORKER} plugins.}
    \vspace{-2mm}
    \label{fig:dccl_simple}
\end{figure}

The network-contention-aware dynamic configuration tuner consists of three primary infrastructure components connected through a pluggable optimizer interface: a Python telemetry library, an NCCL tuner-plugin coordinator, and an online configuration optimizer (Figure~\ref{fig:dccl_simple}). During an NCCL collective call, the tuner:
(a)~captures per-rank GPU statistics and collective call latency via the Python telemetry library (Section~\ref{sec:python_telemetry});
(b)~interposes on the collective call using the NCCL tuner plugin (Section~\ref{sec:plugin_coord}); and
(c)~detects sustained network contention to replace NCCL's default configuration with an optimized selection (Section~\ref{sec:optimizer}).

We propose \emph{DynamICCL} (pronounced as "dynamical"), a network contention aware dynamic and intelligent tuner for collective communication library. Figure~\ref{fig:dccl_detailed} illustrates the complete tuner architecture and its numbered control flow. During the NCCL initialization phase, each rank loads the tuner plugin and connects to its local coordinator \circled{1}, while the GPU monitor initializes alongside the application process \circled{2}, writing telemetry data to shared memory (SHM) via a zero-copy mechanism \circled{3}. When the AI/ML training workload invokes a collective operation \circled{4}, the plugin intercepts the call \circled{5}. The leader rank plugin sends a configuration request to the skeleton coordinator \circled{6}, which performs a constant-time KV cache lookup \circled{7}\circled{8}. The fetched configuration is then propagated across nodes using an inter-node broadcast \circled{9} and relayed locally to all worker plugins on the node \circled{10}. Concurrently, after dispatching the configuration reply, an asynchronous background thread \circled{11} reads the latest telemetry from SHM \circled{12} and feeds it to contention detector agent (LSTM + SPRT) for contention inference. If contention detector agent detects \texttt{SUSTAINED-CONTENTION} \circled{13}, configuration selector agent (D3QN) evaluates the updated network state to select an optimized configuration, writing it back to the KV cache \circled{14} for subsequent collective operations.
 
%%%=======================================================
\subsection{Telemetry and Shared-Memory Pipeline}
\label{sec:python_telemetry}

The telemetry pipeline captures per-collective latency and GPU metrics off the critical path using two concurrent threads (Figure~\ref{fig:python_lib}). The main thread interposes on PyTorch collective calls with high-precision timers to measure per-call execution latency. Concurrently, a background thread samples GPU statistics using the NVML API~\cite{nvml} at 100~ms intervals (a polling frequency bounded by NVML's hardware read latency) collecting Streaming Multiprocessor (SM) utilization, memory utilization, PCIe transmit/receive throughput, and power consumption. Both telemetry streams are written to a per-rank, shared-memory circular buffer using a zero-copy, lock-free design, allowing the coordinator process to asynchronously read the latest sample without lock contention. Because the telemetry library integrates directly at PyTorch's collective communication layer, it remains fully compatible with alternative backend extensions such as RCCL, NCCLX and MSCCL++ without requiring manual application instrumentation.

%%%=======================================================
\subsection{Coordinator Architecture}
\label{sec:plugin_coord}

 %% ----- Communication protocol
NCCL's Tuner Plugin API serves as our tuner's entry point into NCCL, using three hooks from API~v4 to intercept every collective scheduling decision. During initialization, each rank on a node connects to its local coordinator through TCP; the coordinator designates the first connecting local-rank as the \texttt{LEADER} and all remaining local-ranks as \texttt{WORKERS}. Figure~\ref{fig:dccl_simple} illustrates the control flow when a collective operation is intercepted by the tuner's plugin-coordinator architecture. At every collective invocation, the local-rank \texttt{LEADER} transmits the operation metadata $(collType, nBytes)$ to the local coordinator \circled{1}, which fetches the active configuration $(A, P, nCh, nTh)$ from the optimizer \circled{2}. This configuration is then distributed hierarchically: the Node-0 \texttt{SKELETON} coordinator executes an inter-node broadcast to the \texttt{STUB} coordinators on Nodes 1 through $N-1$ \circled{3}, after which each coordinator performs an intra-node broadcast to deliver the configuration to its local rank plugins \circled{4}. Accessing the thread-count parameter ($nTh$) requires a one-time lightweight patch to NCCL's plugin interface, provided in the dynamic tuner's source repository. Encapsulating the optimizer inside the coordinator eliminates all PyTorch/C++ (\texttt{libtorch}) dependencies from the plugin, keeping it a portable, zero-dependency C/C++ library compatible across arbitrary NCCL deployments.

The coordinator runs as a standalone process on each node, compatible with both SLURM-managed clusters and bare-metal environments. Under its \texttt{SKELETON}/\texttt{STUB} topology, one designated node executes in \texttt{SKELETON} mode (hosting the primary optimizer), while all other nodes run as lightweight \texttt{STUB} processes. The \texttt{SKELETON} distributes updated configurations across node boundaries using \texttt{MPI\_Bcast}; each \texttt{STUB} then relays the configuration locally to its per-rank plugins. This two-tier hierarchy restricts cross-node control traffic to a single broadcast per reconfiguration event, successfully decoupling coordination overhead from NCCL's rank count to number of participating nodes.

%%%=======================================================
\subsection{Optimizer Design}
\label{sec:optimizer}

The optimizer interacts with the coordinator using a non-blocking, two-function interface. On each collective call, \texttt{process()} performs a constant-time KV-cache lookup keyed by $(collType, nBytes)$ and immediately returns the active configuration. Once the reply is dispatched, \texttt{observe()} snapshots the latest telemetry and signals the asynchronous optimizer thread (Algorithm~\ref{alg:background}).

The optimizer thread executes continuously in the background (Algorithm~\ref{alg:background}, line~1). Upon awakening, it captures a snapshot of current collective metadata, active configuration parameters, and GPU metrics (line~2) and unconditionally invokes contention detector (line~3). At iteration~0, the inference output is discarded to eliminate NCCL cold-start latency artifacts while preserving the LSTM hidden state $\mathbf{h}_0$ (lines~4--5). During iterations 1--8, contention detector agent accumulates temporal context across collective calls without emitting classification decisions (line~14). From iteration~8 onward, if a cool-down period remains active from a prior reconfiguration, it is decremented (lines~8--9); otherwise, an emission of \texttt{SUSTAINED-CONTENTION} triggers configuration selector (D3QN) agent. The configuration selected by the configuration selector agent is written back to the KV cache, and a cool-down counter of \texttt{COOLDOWN\_TICKS}~=~3 is set (lines~10--12), preventing premature re-triggering before the newly applied configuration has stabilized across sufficient iterations.

\begin{algorithm}[t]
\caption{Asynchronous Optimizer Thread}
\label{alg:background}
\begin{algorithmic}[1]

\small
\Statex \textbf{Runs asynchronously} during forward/backward pass
\Statex \textbf{Constants:} COOLDOWN\_TICKS = 3

\While{running}
    \State Wait on condition variable \Comment{woken by observe()}
    \State Snapshot (info, cfg, stats) under mutex
    \State action $\gets$ \textit{contention\_detector}(info, cfg, stats) \Comment runs always
    \If{global\_tick = 0} \Comment{NCCL cold-start overhead}
        \State Discard contention\_detector output, keep LSTM state
    \ElsIf{tick $\geq$ 8} \Comment{Normal operation after warmup}
        \If{cooldown $>$ 0}
            \State cooldown $\gets$ cooldown $-$ 1 \Comment{New config settling}
        \ElsIf{action = SUSTAINED}
            \State Run \textit{configuration\_selector}, update KV\_Cache
            \State cooldown $\gets$ COOLDOWN\_TICKS
        \EndIf
    \EndIf
    \Comment{Ticks 1--8: contention\_detector builds LSTM context only}
    \State tick $\gets$ tick + 1
\EndWhile

\end{algorithmic}
\end{algorithm}

%%==============================================
\subsubsection{Contention Detection}
\label{sec:agent1}

At each iteration, contention detector agent (detailed in Algorithm~\ref{alg:agent1}) constructs a 15-dimensional input vector by concatenating 11 scalar features, including  contention ratio '$\contentionratio$' (Equation~\ref{eq:contention_ratio}), log-scaled topology parameters, normalized GPU statistics, and a configuration-change flag—with a 4-dimensional learned embedding of the collective operation type (lines~1--4). This vector is processed by an LSTM layer with a 128-dimensional hidden space, where the recurrent hidden and cell states $(\mathbf{h}_t, \mathbf{c}_t)$ persist across consecutive iterations (line~5). Passing the LSTM hidden state through a feedforward classification head yields the sustained contention probability $p_{\text{sustained}}$ (line~6), which is subsequently evaluated against the Sequential Probability Ratio Test (SPRT) thresholds from Equation~\ref{eq:sprt} (lines~7--12). With a compact memory footprint of approximately 358~KB, contention detector agent completes inference in sub-millisecond time on a CPU core.

\begin{algorithm}[t]
\caption{Network Contention Detector Agent}
\label{alg:agent1}
\begin{algorithmic}[1]

\small
\Statex \textbf{Input:} collInfo (collType, nRanks, nNodes, nBytes), cfg, stats
\Statex \textbf{Output:} action (WAIT / SPIKE / SUSTAINED)
\Statex \textbf{Persistent:} LSTM hidden state (h, c)

\State r $\gets$ latency / baseline \Comment{Congestion ratio}
\State scalars $\gets$ [r, CollectiveInfo, GpuStats, cfg\_changed] \Comment{11-dim}
\State emb $\gets$ \textit{Embed}(collType) \Comment{5 $\to$ 4-dim}
\State x $\gets$ \textit{Concat}(scalars, emb) \Comment{15-dim}
\State out, (h, c) $\gets$ \textit{LSTM}(x, h\_prev, c\_prev) \Comment{hidden = 128}
% \State p $\gets$ Sigmoid(Linear(ReLU(Linear(ReLU(Linear(out))))))
\State p $\gets$ \textit{FeedForwardNet}(out)\Comment{128 $\to$ 256 $\to$ 128 $\to$ 1}
\If{p $>$ 0.95} \Comment{SPRT upper threshold}
    \State \Return SUSTAINED
\ElsIf{p $<$ 0.05} \Comment{SPRT lower threshold}
    \State \Return SPIKE
\Else
    \State \Return WAIT
\EndIf

\end{algorithmic}
\end{algorithm}

% ------------------------------------------------------------
\begin{figure}[b]
    \centering
    \includegraphics[width=1.0\columnwidth]{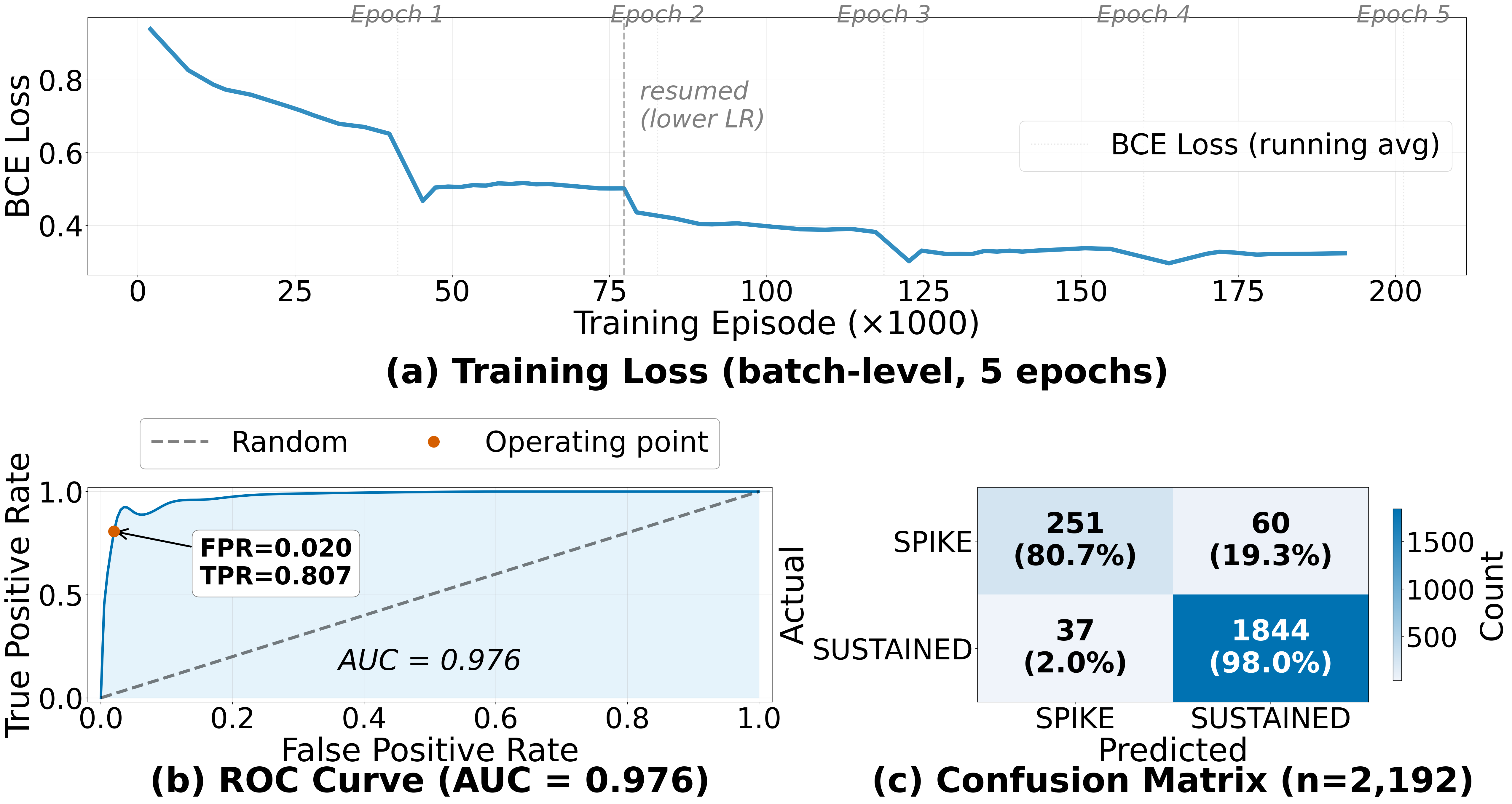}
    \caption{Contention detector agent training evaluation}
    \label{fig:agent1_training}
\end{figure}
% -----------------------------------------------------------

Contention detector agent is trained offline with class-balanced sampling at 17,400 episodes per class per epoch to prevent the minority SPIKE class from being too underrepresented.
Figure~\ref{fig:agent1_training} shows BCE loss stabilizing by epoch~3 (a), AUC=0.976 (b), and 97.5\% sustained recall with less than 5\% false-positive rate on held-out episodes (c).
Detection speed adapts to severity: approximately 4 iterations at 80\% contention and 9 at 50\%.

%%==============================================
\subsubsection{Configuration Selection}
\label{sec:agent2}

Configuration selector agent (detailed in Algorithm~\ref{alg:agent2}) is invoked exclusively upon a \texttt{SUSTAINED-CONTENTION} emission from contention detector. It first constructs a contention vector by encoding contention direction, magnitude, and confidence from contention detector's output (line~1), assembles a 153-dimensional state vector combining current system telemetry with the recurrent LSTM hidden state $\mathbf{h}_t$ (lines~2--3), and forwards it through the trunk network (line~4). The dueling streams evaluate a state-value estimate $V(s)$ and per-action advantages $A(s, a)$ (lines~5--6), which are aggregated to compute Q-values (lines~7--9). The greedy action $a^* = \arg\max_a Q(s, a)$ is then decoded into a valid configuration tuple (lines~10--14) and written to the KV cache for the subsequent collective operation (line~15). With a compact memory footprint of approximately 168~KB, configuration selector agent executes CPU inference in under 2~ms.

\begin{algorithm}[t]
\caption{Configuration Selector Agent}
\label{alg:agent2}
\begin{algorithmic}[1]

\small
\Statex \textbf{Input:} LSTM hidden state h, info, cfg, stats
\Statex \textbf{Output:} Config (A, P, nCh, nTh)
\Statex \textbf{Action space:} 2 Algo $\times$ 2 Proto $\times$ 8 Ch $\times$ 3 Th = 96
\State cong $\gets$ \textit{CongestionVector}(action, r, p\_sustained)
\Statex \Comment{direction $\in$ [-1,1], magnitude $\in$ [0,3], confidence $\in$ [0,1]}
\State state $\gets$ [Congestion, collInfo,last\_config, GPU\_stats]
\State x $\gets$ [state, LSTM\_history]\Comment{input vector size 153}
\State z = \textit{DQN}(x)\Comment{153 $\to$ 256 $\to$ 128} 
\State V $\gets$ \textit{FeedForwardNet}(z) \Comment{Value: 128 $\to$ 64 $\to$ 1}
\State A $\gets$ \textit{FeedForwardNet}(z) \Comment{Advantage: 128 $\to$ 64 $\to$ 96}
\For {a$\in$Action Space}
    \State Q(a) $\gets$ V + A(a) $-$ mean(A)\Comment{Dueling aggregation}
\EndFor
\State idx $\gets$ argmax Q(a) \Comment{Greedy selection}
\State nTh $\gets$ \{128, 256, 512\}[idx mod 3]
\State nCh $\gets$ (idx / 3) mod 8 + 1
\State P $\gets$ \{LL, Simple\}[(idx / 24) mod 2]
\State A $\gets$ \{Tree, Ring\}[idx / 48]
\State Update KV\_Cache[(collType, nBytes)] $\gets$ (A, P, nCh, nTh)
\State \Return (A, P, nCh, nTh)

\end{algorithmic}
\end{algorithm}

%------
\begin{figure}[t]
    \centering
    \includegraphics[width=1.0\columnwidth]{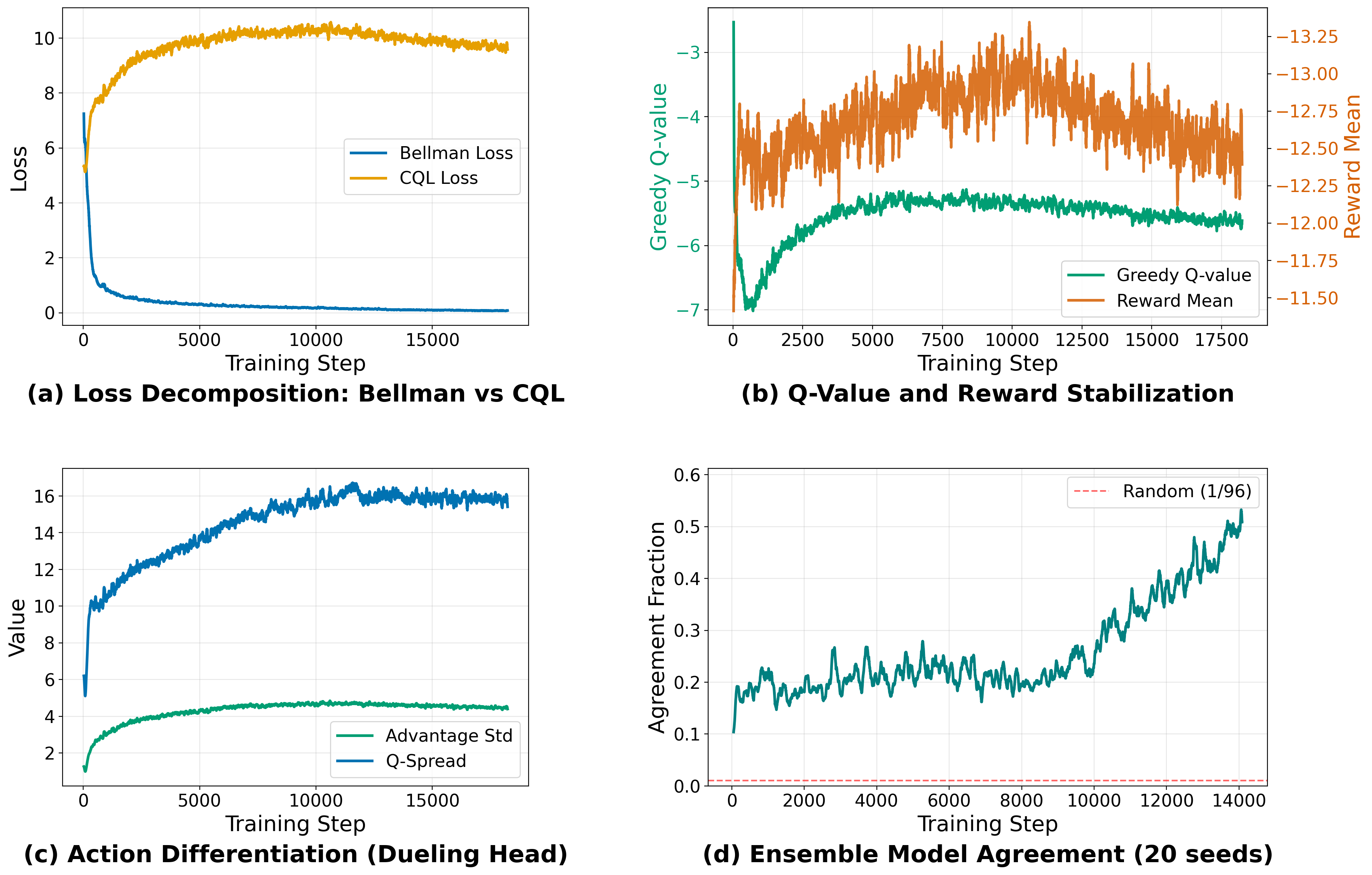}
    \caption{Configuration selector agent training evaluation}
    \vspace{-2mm}
    \label{fig:agent2_training}
\end{figure}
%------

COnfigruation selector agent is trained offline using Dueling Double DQN with Conservative Q-Learning~\cite{cql} regularization ($\alpha_{\mathrm{CQL}} = 0.001$), optimizing the reward function and loss formulation defined in Equations~\ref{eq:reward} and~\ref{eq:cql}. Figure~\ref{fig:agent2_training} confirms empirical convergence within 10K training steps, demonstrating Q-value stability, clear dueling-head differentiation, and strong ensemble agreement across 20 independent random seeds compared to the baseline.

%%==============================================
\subsection{Implementation}
\label{sec:implementation}

The network-contention-aware dynamic configuration tuner is implemented in approximately 4,250 lines of C++17 across four core modules: the coordinator, tuner plugin, online optimizer, and TCP communication utilities. An additional 610 lines of Python comprise the telemetry library. The system is built using GNU Make with \texttt{-std=c++17 -O2 -fPIC}, eliminating external build dependencies like CMake. Core runtime dependencies are restricted to POSIX shared memory, POSIX-threads, a CPU-only PyTorch build (\texttt{libtorch}~2.7.1), and MPI for multi-node deployments. To maintain isolation, the build vendors only the NCCL Tuner Plugin ABI headers without linking against NCCL at compile time.

\textbf{Telemetry library.}
The Python telemetry library runs as a process alongside each GPU rank, capturing hardware metrics from the NVML API and logging per-call collective latencies into a per-rank POSIX shared-memory region. Each shared-memory buffer contains 1,024 slots of 64~bytes each, prepended by a 64-byte header maintaining a monotonically incrementing write index (totaling 65,600~bytes per rank). The coordinator reads the latest sample in $\mathcal{O}(1)$ time by evaluating \texttt{(write\_index - 1) \% 1024}, enabling non-blocking, zero-copy reads without lock contention.

\textbf{Tuner plugin.}
The plugin is compiled as a shared library loaded by NCCL at runtime using the \texttt{NCCL\_TUNER\_PLUGIN} environment variable. It implements the NCCL Tuner Plugin ABI~v4.1 interface across three hooks: \texttt{init}, \texttt{getCollInfo}, and \texttt{destroy}. By isolating machine learning dependencies to the coordinator binary, the plugin remains a compact C++ component of approximately 460 lines. When the coordinator emits a \texttt{REPLY\_BYPASS} signal, the plugin returns \texttt{ncclSuccess} without modifying NCCL's cost tables, delegating configuration selection directly to NCCL's built-in heuristics. Furthermore, if the coordinator fails to initialize, NCCL transparently falls back to its default runtime behavior without interrupting the workload.

% ------------------------------------------------------------
\begin{figure}[t]
    \centering
    \includegraphics[width=0.7\columnwidth, height=0.8\columnwidth]{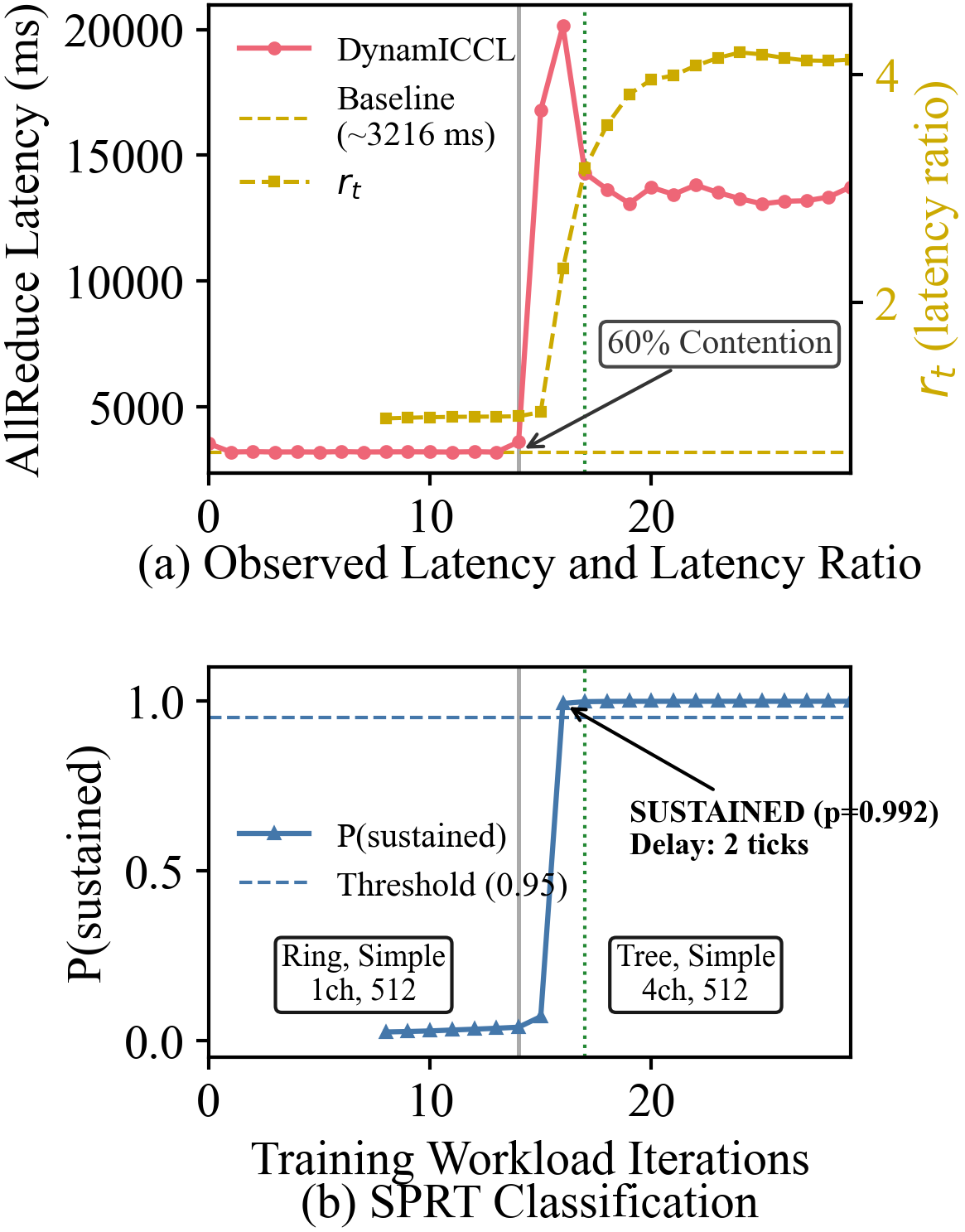}
    \caption{Contention Detection Timeline}
    \vspace{-2mm}
    \label{eval:CD_timeline}
\end{figure}

\textbf{Coordinator.}
The coordinator is an MPI application executed using \texttt{mpirun}, deploying exactly one MPI process (rank) per cluster node hosting the NCCL workload. The \texttt{SKELETON} process (MPI rank~0) hosts the online optimizer and handles configuration decisions. \texttt{STUB} processes (MPI ranks~1 through $N-1$) receive decisions from the \texttt{SKELETON} through \texttt{MPI\_Bcast} using a compact message structure comprising of five \texttt{uint32\_t} fields, further relaying parameters to node-local per-rank plugins over TCP on \texttt{localhost}. This design scales internode coordination overhead with node count rather than rank count (Section~\ref{sec:cord_plugin_why}). In production, coordinators are launched prior to the application using \texttt{mpirun}, followed by the training job via \texttt{torchrun}. The coordinator is validated across SLURM-managed clusters and bare-metal environments. To simplify local testing and single-node development, a TCP-only coordinator variant is also provided that operates without requiring an MPI runtime.

\textbf{Optimizer.}
Both inference models are serialized as TorchScript binaries and loaded in C++ using \texttt{torch::jit::load()}. The recurrent LSTM hidden and cell states $(\mathbf{h}_t, \mathbf{c}_t)$ of dimension $1 \times 1 \times 128$ are maintained as C++ \texttt{torch::Tensor} instances across collective iterations. Inference executes in FP32 on CPU without quantization, completing in under 3.6~ms combined for both agents. Under the \texttt{REPLY\_BYPASS} path—active during initial warmup as well as on \texttt{SPIKE} or \texttt{WAIT} emissions, configuration selector agent is bypassed entirely, incurring zero computational overhead. 

Source code for the network-contention-aware dynamic configuration tuner  will be made publicly available on GitHub upon publication.

\section{Evaluation}
\label{sec:eval}

We evaluate the dynamic tuner on two platforms selected for complementary evaluation objectives: Chameleon Cloud, which provides isolated bare-metal nodes across heterogeneous GPU generations (A100, V100, and P100) for controlled microbenchmark experiments, and the Delta cluster, which we use to validate scaling behavior. We compare three systems built against identical NCCL v2.29.2+cuda12.6 binaries: the default NCCL heuristic, AutoCCL's coordinate-descent tuner~\cite{autoccl}, and our dynamic tuner. Speedups are computed using a paired per-run methodology: the $N$-th run of a baseline is matched with the $N$-th run of our dynamic tuner, per-pair speedups are computed, and summary statistics are reported over the resulting distribution. We report the mean for microbenchmark results, where isolated bare-metal nodes yield low run-to-run variance, and the median for scaling results, where stochastic background traffic on the shared cluster introduces heavy-tailed variance that would otherwise skew the mean.

%%==========================================================================
\subsection{Microbenchmark Results}
\label{sec:eval_micro}

The microbenchmark testbed comprises four nodes (two with dual P100-SXM2 GPUs and two with dual P100-PCIe GPUs), interconnected via a 10~Gbps full-duplex network. Collective operations on 2~GB \texttt{float32} tensors serve as the baseline workload, supplemented by tensor-size sweeps ranging from 256~MB to 4~GB. Network contention is injected using symmetric BBR \textit{iperf3} background streams that consume 50\%--90\% of link capacity across both inter-node pairs, with precise onset control.

% ------------------------------------------------------------
\begin{figure}[t]
    \centering
    \includegraphics[width=1.0\columnwidth, height=0.6\columnwidth]{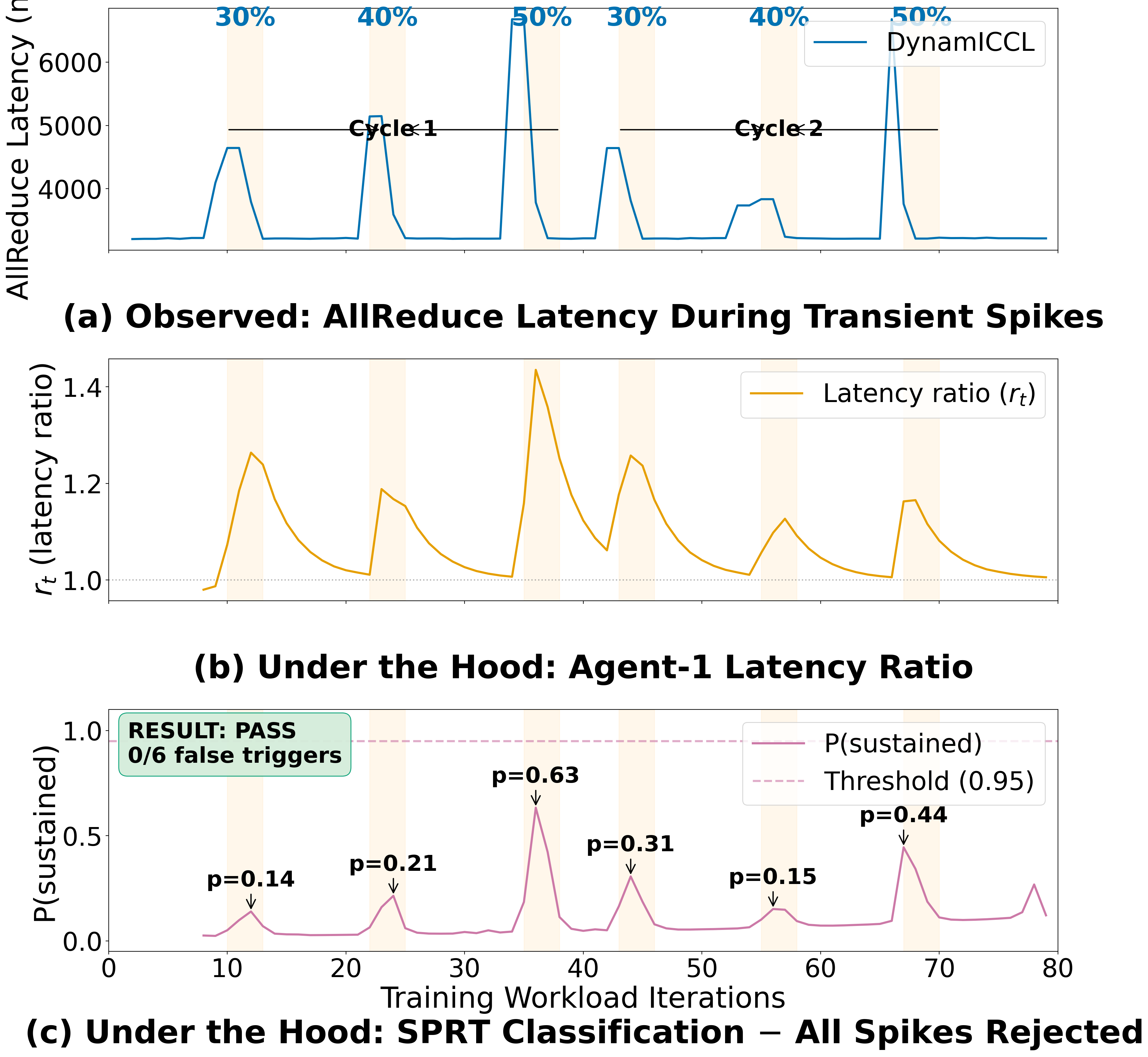}
    \caption{Transient spike rejection under short contention bursts.}
    \vspace{-2mm}
    \label{eval:spike}
\end{figure}
% ------------------------------------------------------------
\begin{figure}[b]
    \centering
    \includegraphics[width=1.0\columnwidth,height=0.4\columnwidth]
    {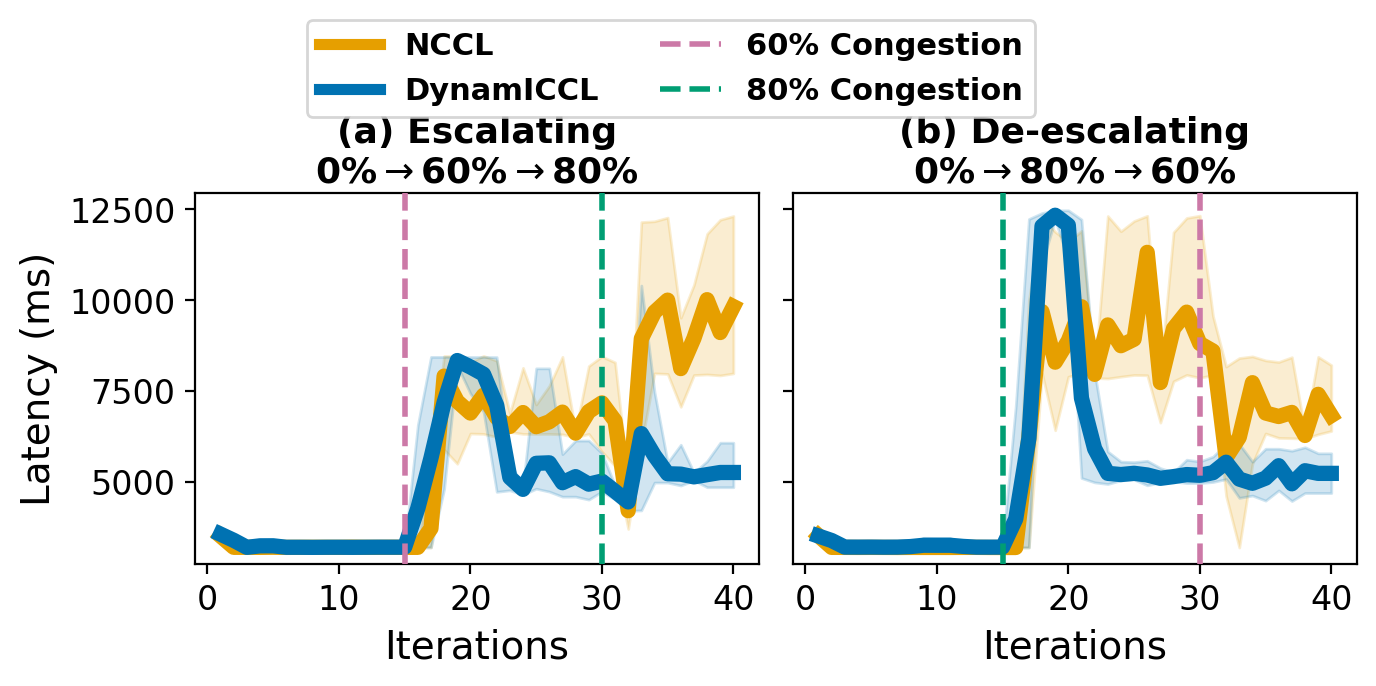}
    \caption{Multi-level contention transition response.}
    \label{eval:multi_levl_cong}
\end{figure}
% ------------------------------------------------------------
\begin{figure}[t]
    \centering
    \includegraphics[width=1.0\columnwidth]{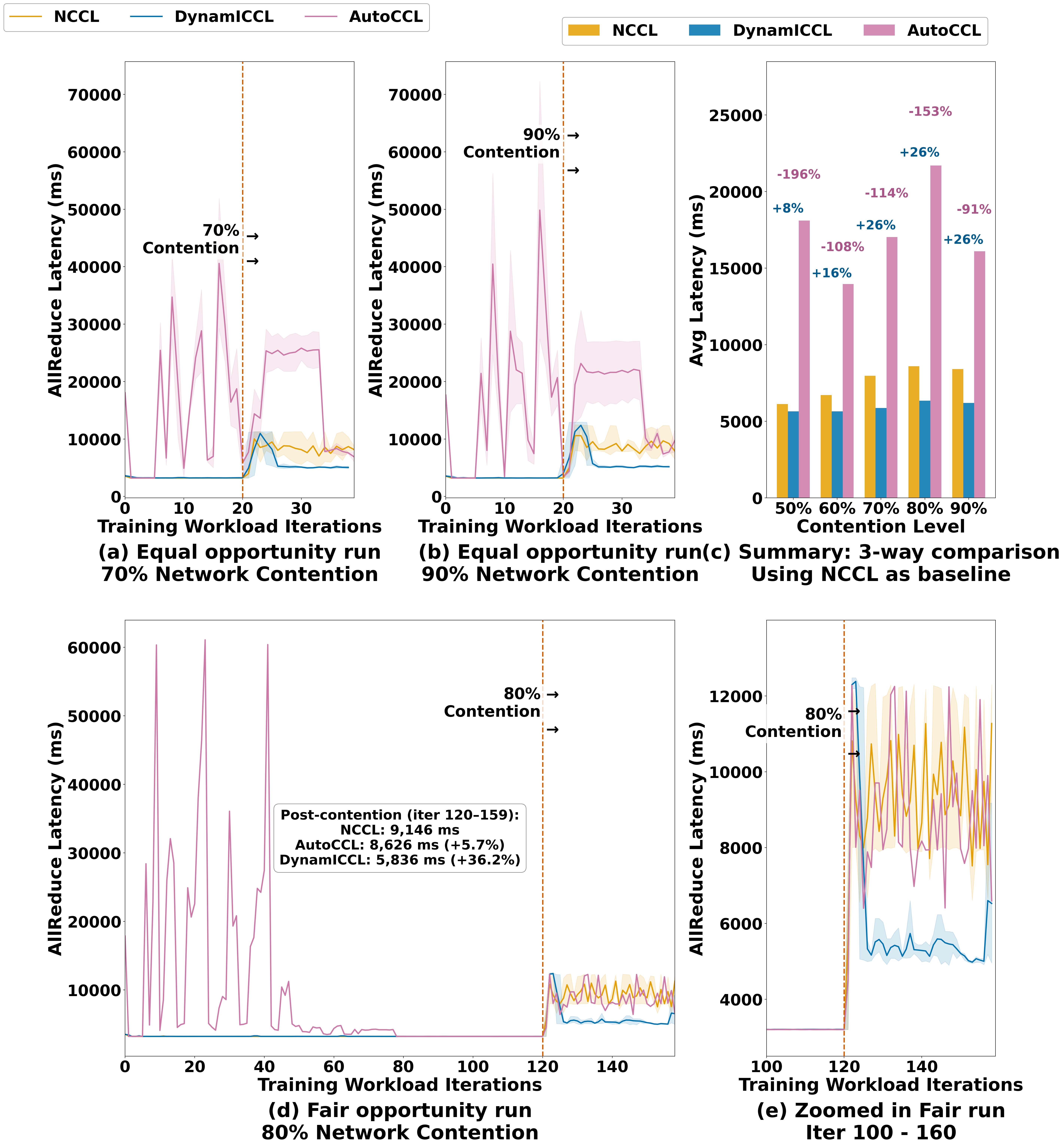}
    \caption{Equal-opportunity and fair-opportunity comparison: NCCL vs.\ AutoCCL vs.\ our dynamic tuner.}
    \vspace{-2mm}
    \label{eval:3way_combined}
\end{figure}
% ------------------------------------------------------------
\begin{figure}[t]
    \centering
    \includegraphics[width=.70\columnwidth, ]{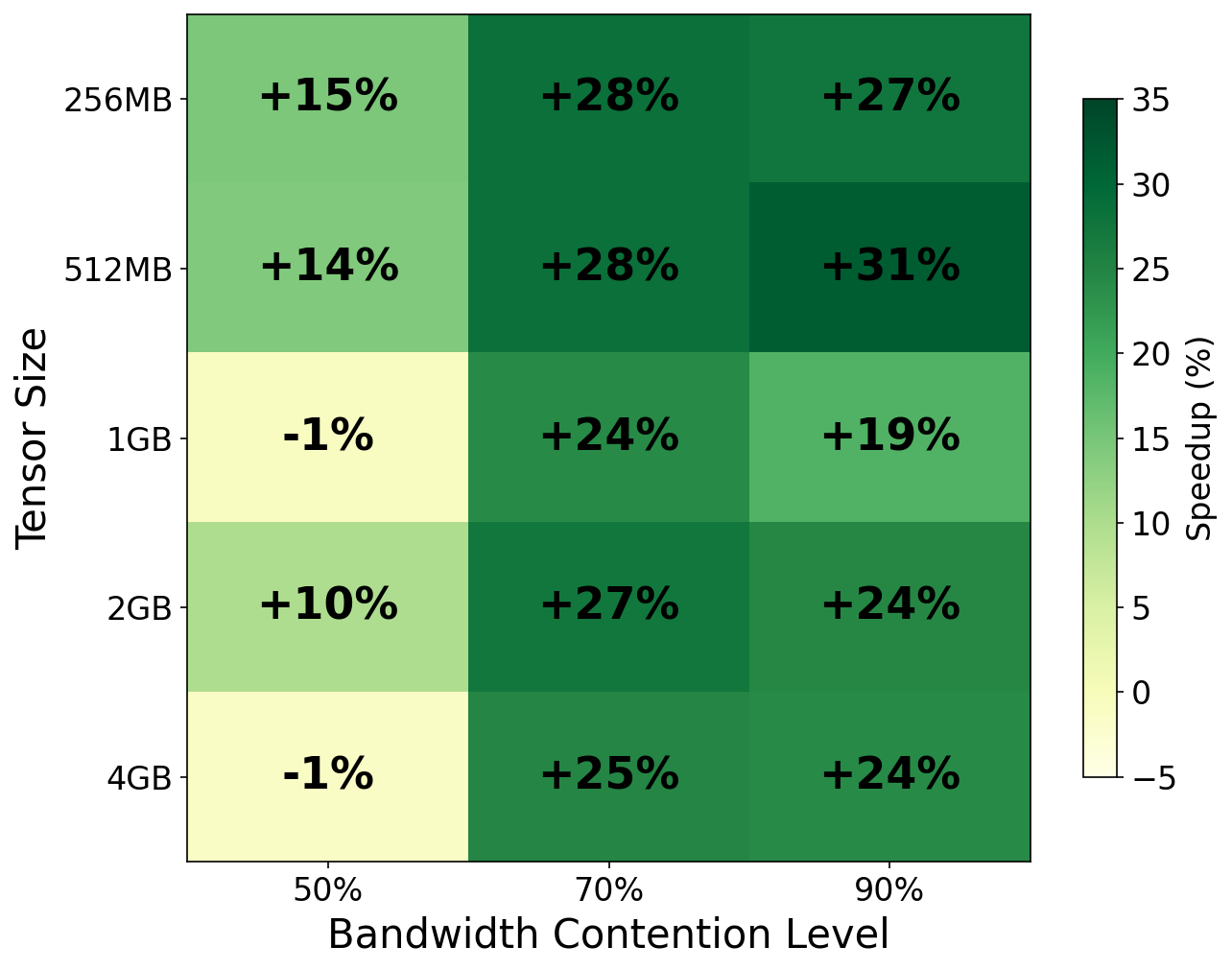}
    \caption{Heatmap of speedup across varying tensor sizes and network contention levels.}    
    \vspace{-2mm}
    \label{eval:heatmap}
\end{figure}
% ------------------------------------------------------------
\begin{figure}[b]
    \centering
    \includegraphics[width=1.0\columnwidth]{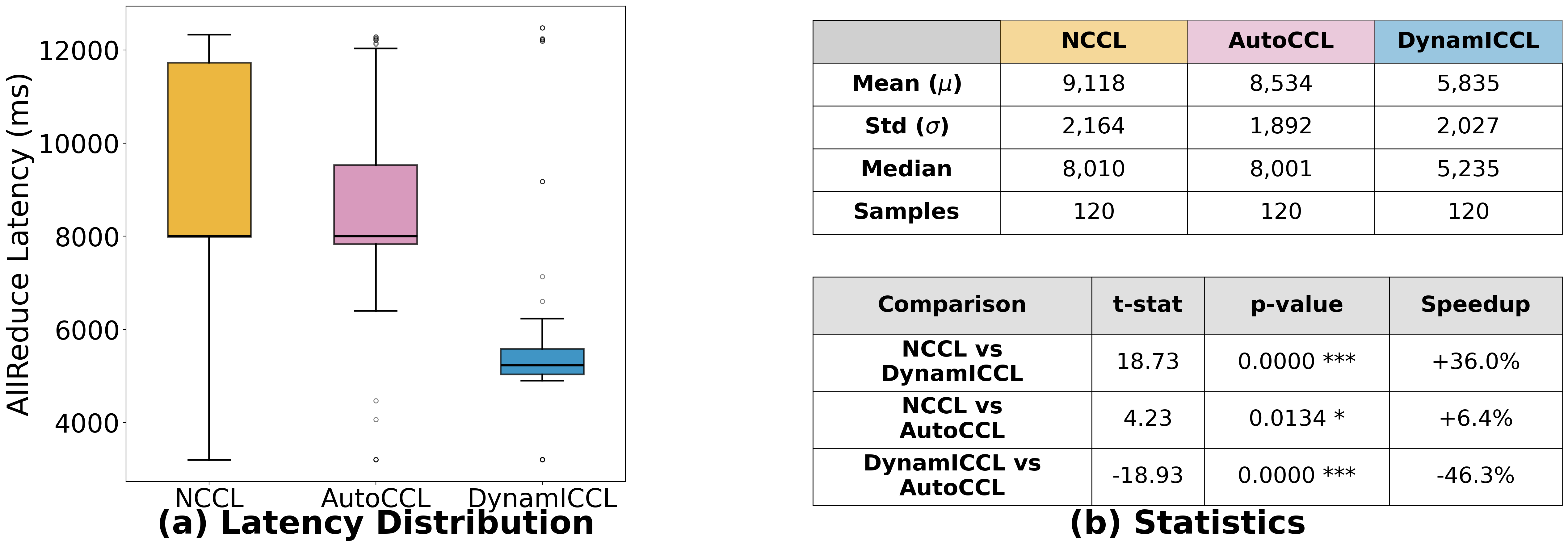}
    \caption{Latency distribution and $t$-test results comparing default NCCL, AutoCCL, and the dynamic tuner.}
    \label{eval:boxplot}
\end{figure}
% ------------------------------------------------------------

\subsubsection{Runtime Contention Detection}
\label{sec:eval_detection}

Contention detector agent's LSTM recurrent network accumulates evidence proportional to contention severity, adjusting its detection rate dynamically. As illustrated in Figure~\ref{eval:CD_timeline}, the agent detects sustained contention in approximately 4 iterations under 80\%--90\% link congestion and 9 iterations at 50\% congestion, with $p_{\text{sustained}}$ crossing the SPRT threshold ($0.95$) proportionally faster under heavier degradation. This adaptive response is a direct benefit of pairing the LSTM classifier head with a Sequential Probability Ratio Test (SPRT) to evaluate decision confidence: unlike fixed-window heuristic detectors, it avoids imposing a rigid detection delay under severe congestion."

Equally critical is the reliable rejection of transient network noise. Figure~\ref{eval:spike} demonstrates that the detector agent correctly classifies all 2-iteration contention spikes (at 30\%, 40\%, and 50\% severity across six total bursts), resulting in zero false reconfigurations and keeping configuration selector agent uninvoked.

To isolate onset timing from adaptation speed, we evaluate contention injected at iterations 10, 20, and 30 under 80\% link congestion. The post-reconfiguration iteration speedup remains consistent at approximately 42\% regardless of onset timing, confirming that the contention detector agent's state progression does not depend on specific temporal iteration indices.

Figure~\ref{eval:multi_levl_cong} illustrates the dynamic tuner handling both escalating ($0\% \to 60\% \to 80\%$) and de-escalating ($0\% \to 80\% \to 60\%$) congestion shifts. In both patterns, the system executes a single, stable configuration switch, holding average latency within 5,000--5,800~ms compared to NCCL's 6,500--8,400~ms without triggering redundant re-optimizations.

%%=====================================================================================
\subsubsection{Performance Analysis}
\label{sec:eval_perf}

We evaluate all systems under two complementary execution regimes. The \emph{equal-opportunity} regime initializes all systems concurrently and measures performance from the first iteration under contention, reflecting realistic cold-start online deployment. The \emph{fair-opportunity} regime grants AutoCCL 100 pre-contention iterations to converge prior to measurement, isolating its converged configuration quality from its online search overhead. Together, these regimes decouple runtime search overhead from asymptotic parameter quality.

Figure~\ref{eval:3way_combined} presents per-iteration latency across contention levels for both regimes. Under \emph{equal-opportunity} evaluation, AutoCCL's live coordinate descent active search increases iteration latency by 87\%--195\% relative to baseline NCCL.
At 80\% contention, baseline NCCL averages 8,673~ms, our dynamic tuner achieves 6,147~ms ($+29.1\%$ speedup), and AutoCCL averages 21,697~ms ($-150\%$ slowing), highlighting the efficiency gap between pre-trained CPU-bound inference and online coordinate search.

Under \emph{fair-opportunity} evaluation, AutoCCL achieves a modest 6.4\% speedup over NCCL, whereas our dynamic tuner achieves a 36\% latency reduction over NCCL and a 32\% reduction over AutoCCL ($p < 0.0001$).
The latency distributions in Figure~\ref{eval:boxplot} confirm these improvements with high statistical significance ($p < 0.0001$). Furthermore, the proposed tuner exhibits significantly lower variance, yielding highly predictable iteration times.
Under clean (uncongested) network conditions, pre-contention latency differs by less than 0.2\% between NCCL and our tuner (3,204~ms vs.\ 3,210~ms), confirming negligible runtime overhead.

Across varying contention levels, the dynamic tuner reduces AllReduce latency by 8.2\% (at 50\% congestion), 19.0\% (70\%), 29.1\% (80\%), and 30.2\% (90\%). The dynamically selected new configuration maintains stable iteration times between 5,500--6,100~ms, while default NCCL degrades from 6,136~ms to 8,673~ms.
Figure~\ref{eval:heatmap} shows speedup heatmaps across five tensor sizes (256~MB to 4~GB) and three contention levels.
Under 70\%--90\% congestion, the tuner delivers consistent 19\%--31\% speedups across all tensor sizes, demonstrating that configuration selector agent's policy captures generalizable network-configuration dynamics rather than overfitting to specific payload volumes.

%%==========================================================================
\begin{figure}[t]
    \begin{subfigure}{0.49\linewidth}
        \includegraphics[width=\linewidth, height=45mm]{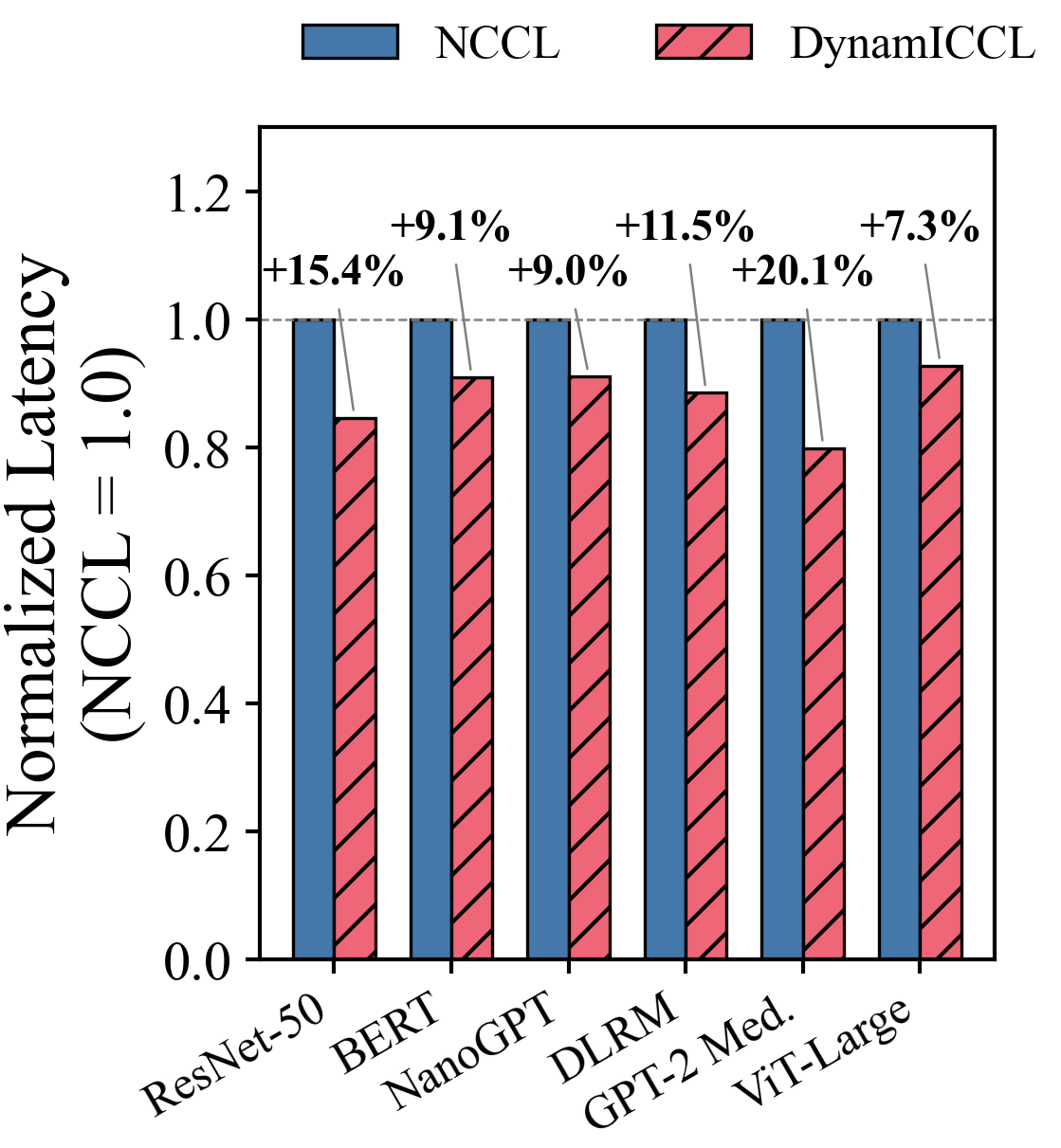}
        \caption{80\% Network contention}
    \end{subfigure}
    \begin{subfigure}{0.49\linewidth}
        \includegraphics[width=\linewidth,height=45mm]{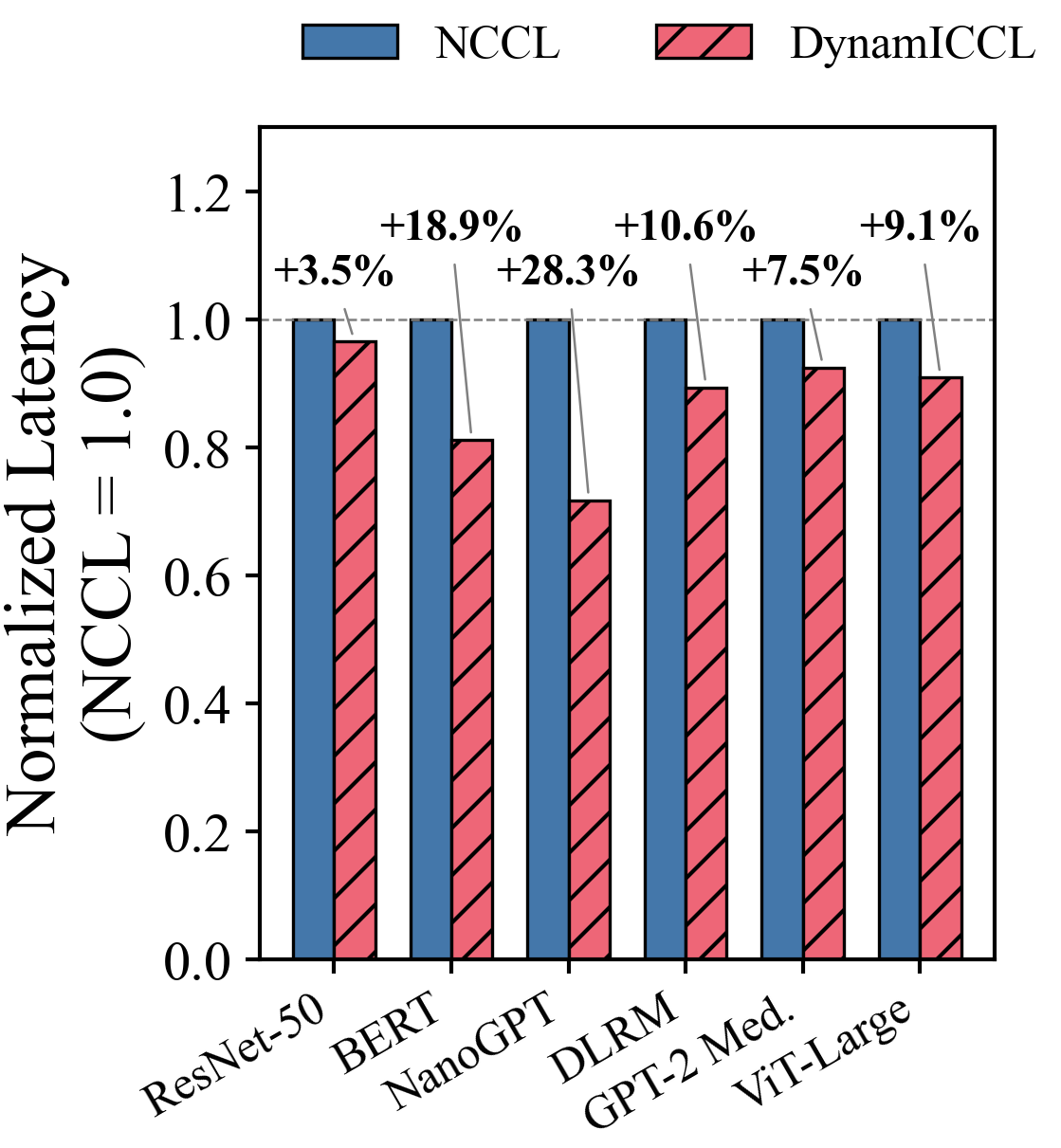}
        \caption{100\% Network contention}
    \end{subfigure}
    \caption{Performance improvement of our dynamic tuner across real AI workloads under network contention.}
    \vspace{-2mm}
    \label{eval:ai_workloads}
\end{figure}

\subsection{AI Workload Analysis}
\label{sec:eval_ai}

We evaluate the dynamic tuner across six representative AI training workloads—ResNet-50, BERT, NanoGPT, DLRM, GPT-2 Medium, and ViT-Large—under 80\% and 100\% network contention.
As shown in Figure~\ref{eval:ai_workloads}, the proposed tuner yields positive median speedups across all 12 workload-contention configurations, ranging from 7\% to 28\%.
At 80\% contention, the average speedup reaches 12.06\%; at near 100\% contention, it increases to 12.98\%, with BERT and NanoGPT achieving median speedups of 18.9\% and 28.3\%, respectively.
These higher gains stem from the large per-iteration AllReduce volumes in transformer architectures, which exacerbate default NCCL routing penalties under congestion.
Conversely, lighter workloads like ResNet-50 and DLRM invoke smaller collectives with lower absolute communication overhead, leaving less room for optimization.
The consistent performance gains across diverse architectures—spanning CNNs (ResNet-50), Transformers (BERT, GPT-2 Medium, ViT-Large), and Recommendation models (DLRM)—confirm that our contention-aware configuration strategy generalizes across varied collective communication patterns.

%%==========================================================================
\subsection{Scaling Analysis}
\label{sec:eval_scaling}

We evaluate system scaling on the Delta cluster (A100 GPUs with Slingshot-11 interconnect) using data-parallel training of 'GPT-2 Medium' model under real background contention. To isolate how contention sensitivity changes with cluster topology, we evaluate both density scaling (1 to 4 GPUs per node across a fixed 4-node cluster) and node scaling (1 GPU per node scaled out from 4 to 24 nodes). This dual-axis design captures the underlying topology dynamics: inter-node traffic exposed to network congestion scales directly with total node count, whereas higher per-node rank densities shift communication locally to intra-node interconnects, reducing exposure to external link contention.

\begin{figure}[t]
    \begin{subfigure}{0.45\linewidth}
        \includegraphics[width=\linewidth, height=45mm]{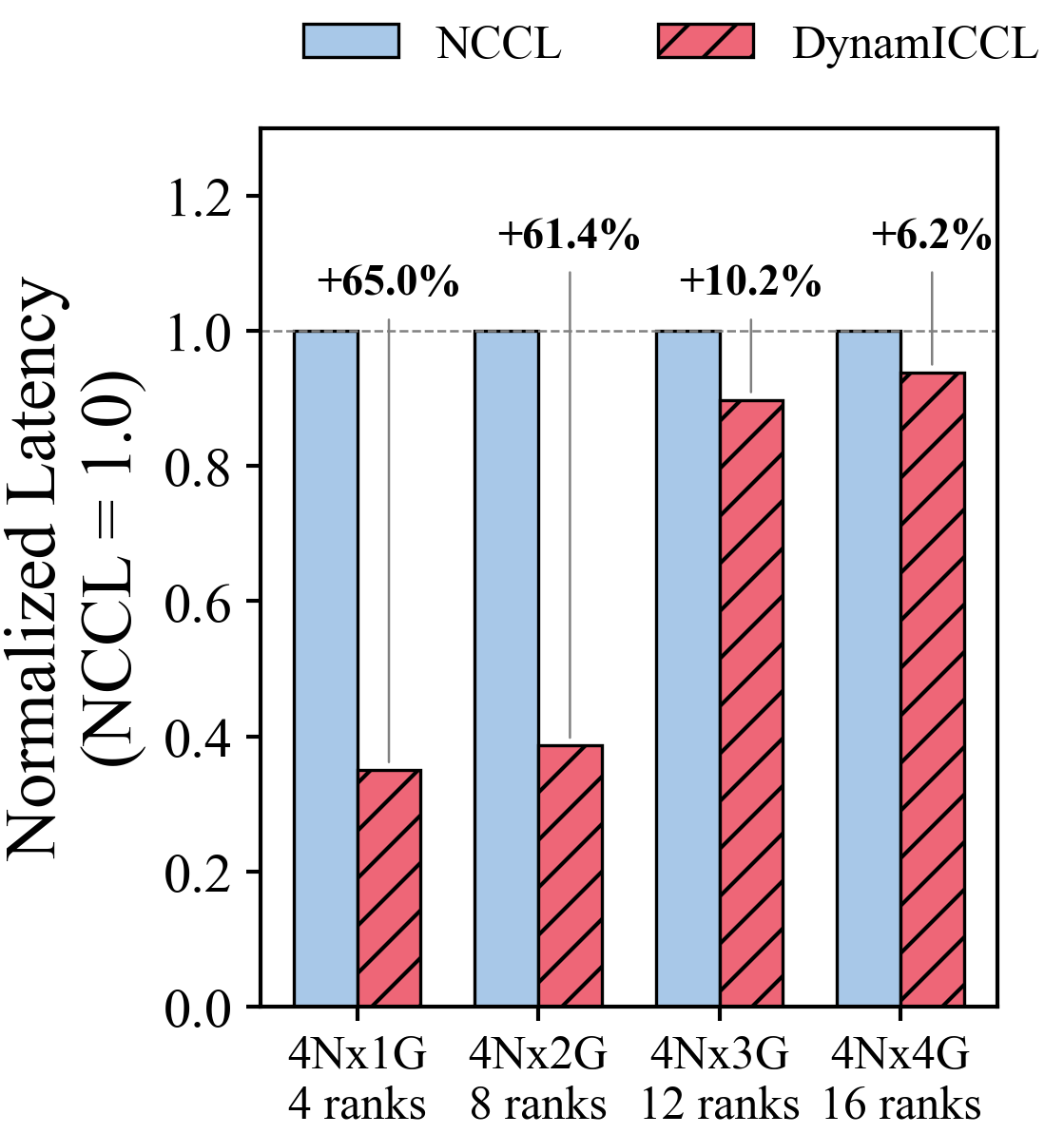}
        \caption{Density scaling}
    \end{subfigure}
    \begin{subfigure}{0.54\linewidth}
        \includegraphics[width=\linewidth, height=45mm]{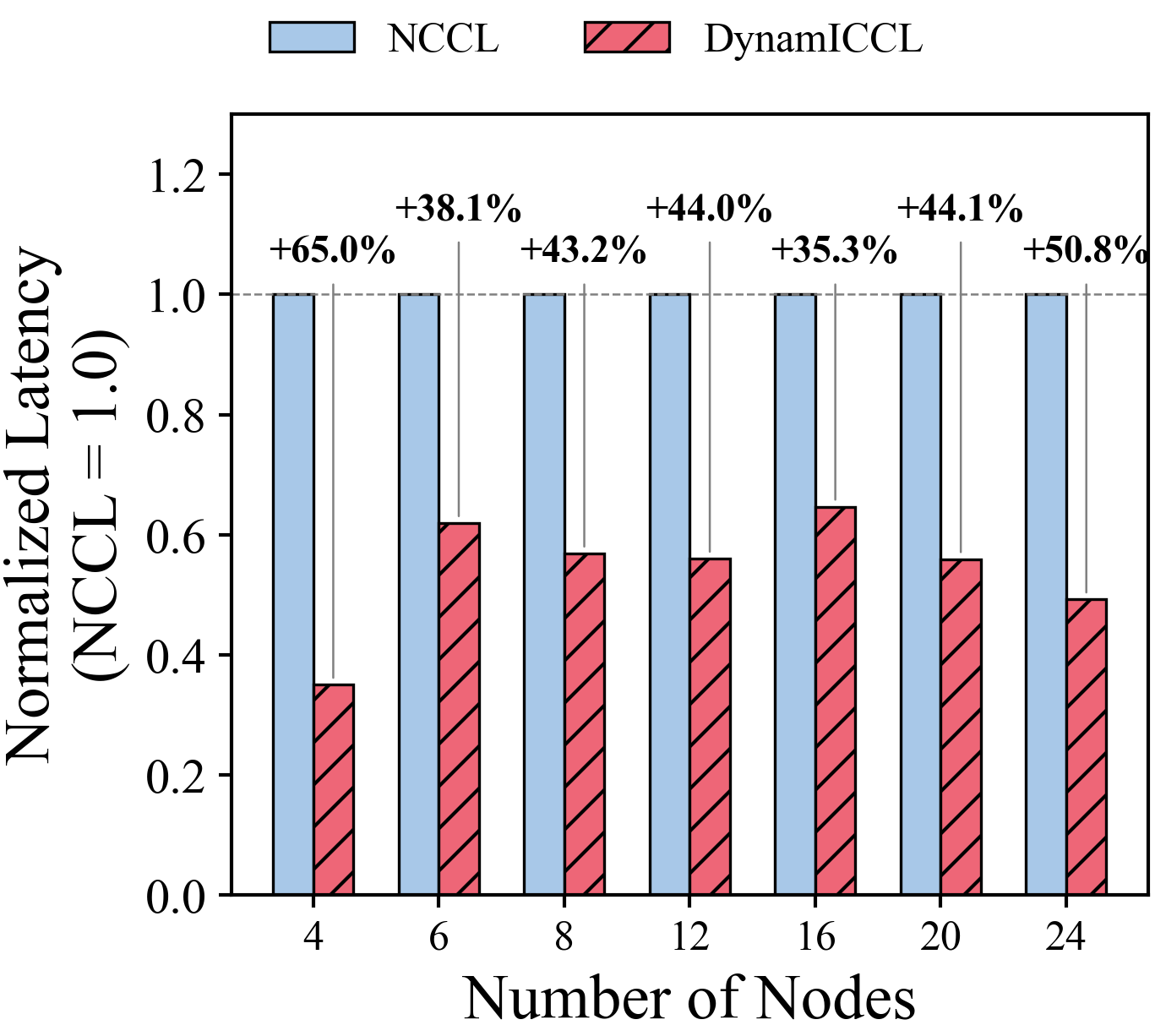}
        \caption{Node scaling}
    \end{subfigure}
    \caption{Scaling behavior of the dynamic tuner for GPT-2 Medium model training on the Delta cluster (A100 GPUs with Slingshot-11).}
    \vspace{-2mm}
    \label{eval:scaling}
\end{figure}

As shown in Figure~\ref{eval:scaling}(a), density scaling done by fixing number of nodes while increasing  number of NCCL ranks yields speedups of 6.2\%--65.0\% (+65.0\% at 4 ranks, +61.4\% at 8 ranks, +10.2\% at 12 ranks, and +6.2\% at 16 ranks), tapering as per-node rank density increases. At higher densities, additional intra-node ranks contend for shared NIC bandwidth, reducing per-rank link headroom and compressing the headroom within which the dynamic tuner can identify a better configuration. Conversely, Figure~\ref{eval:scaling}(b) shows that node scaling maintains consistent speedups between 35.3\% and 65.0\% across 4 to 24 nodes (averaging over +45\% speedup). This confirms that as inter-node communication dominates, the coordinator hierarchy scales efficiently to mitigate the primary network bottleneck regardless of cluster scale. Across all scaling configurations, pre-contention system overhead remains below 3\%.

%% ====================================================
\subsection{Agent Configuration Selection Analysis}
\label{sec:eval_explainability}

A central question in learning-based system optimization models is whether the control policy actively leverages high-dimensional runtime telemetry or merely relies on static workload features. To evaluate whether configuration selector agent's network-awareness mechanism actively drives configuration selection, we conduct a feature sensitivity and gradient attribution analysis across 800 realistic execution states (spanning 5 workload contexts across 160 temporal ticks of a real-world contention trace, with congestion onset at tick 120).

\begin{figure}[t]
    \begin{subfigure}{0.49\linewidth}
        \includegraphics[width=\linewidth, height=45mm]{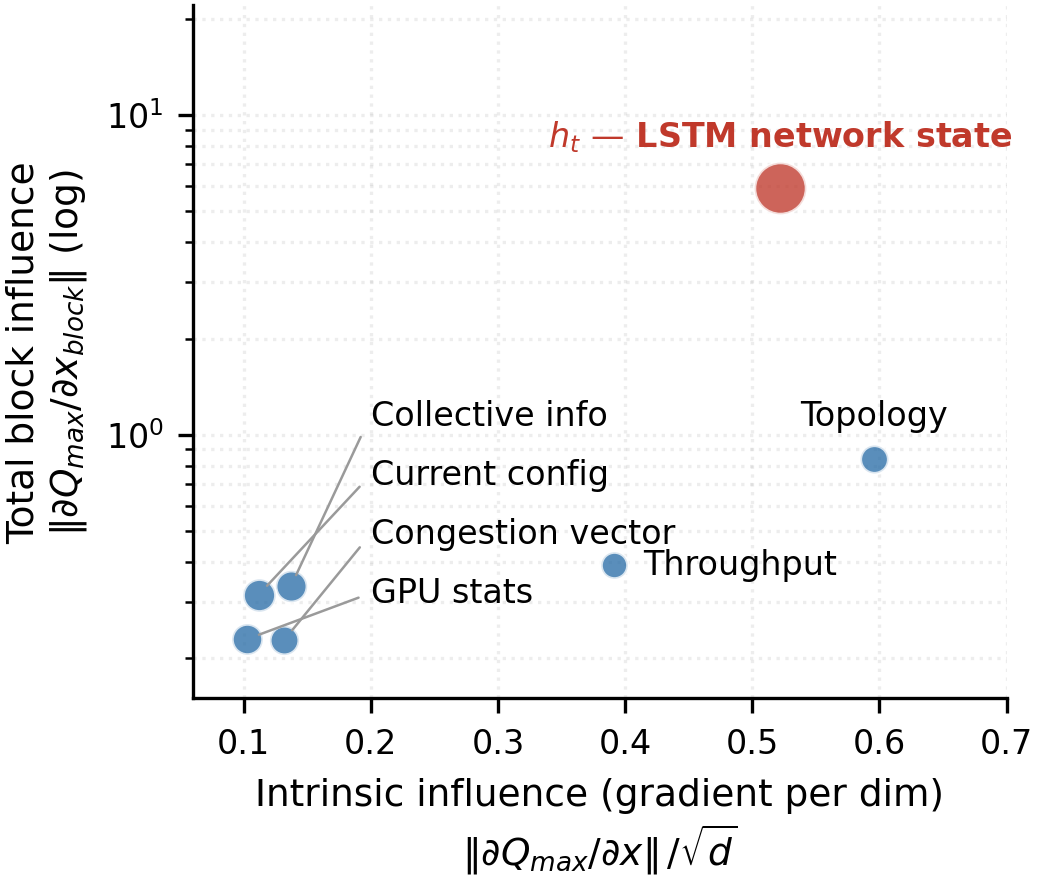}
        \caption{Intrinsic vs. total influence}
    \end{subfigure}
    \begin{subfigure}{0.49\linewidth}
        \includegraphics[width=\linewidth, height=45mm]{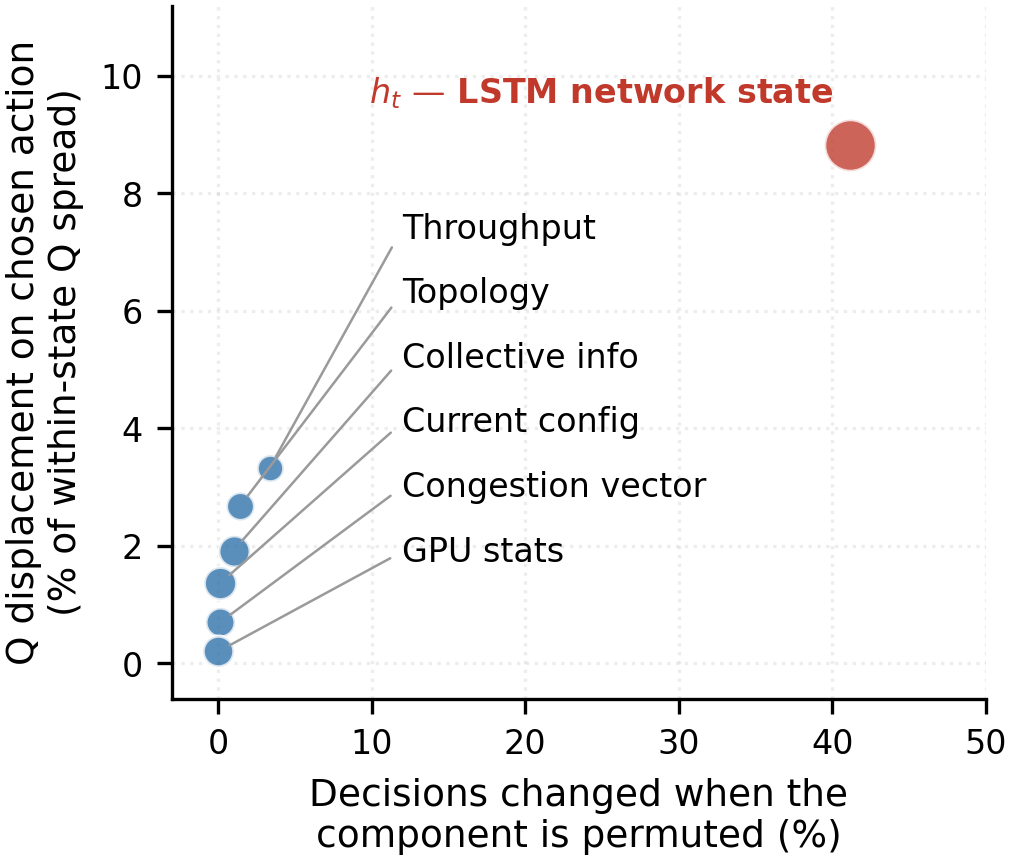}
        \caption{Behavioral impact}
    \end{subfigure}
    \caption{Behavioral impact: share of decisions changed and Q-value displacement when each component is permuted.}
    \vspace{-2mm}
    \label{eval:state_importance}
\end{figure}

We evaluate feature importance across two complementary axes: intrinsic gradient sensitivity ($\|\nabla_{\mathbf{x}} Q\| / \sqrt{d}$) to measure input attention, and decision permutation testing to measure behavioral impact ($Q$-value displacement and percentage of flipped configuration decisions). As shown in Figure~\ref{eval:state_importance}, the results confirm that temporal network state is the primary driver of agent behavior:

\begin{itemize}
    \item \textbf{Intrinsic vs.\ Total Influence} (Figure~\ref{eval:state_importance}--a): To determine if configuration selection  depends on the recurrent network state embedding ($h_t$), we measure output gradient variance relative to input perturbations. Feature-for-feature, $h_t$ demonstrates intrinsic per-dimension sensitivity ($0.522$) comparable to the most dominant static input (topology at $0.596$), outperforming all other state features by $3\times$ to $5\times$. Across its 128 dimensions, $h_t$ accounts for approximately 97\% of total policy sensitivity. Crucially, Figure~\ref{eval:state_importance}(a) resolves the dimensionality question: $h_t$ does not appear dominant merely due to vector size, as each constituent dimension carries substantial individual weight.
    \item \textbf{Behavioral Impact }(Figure~\ref{eval:state_importance}--b): High gradient sensitivity is only meaningful if it alters control outputs. By permuting individual input components while holding others fixed, we isolate their direct effect on parameter selection. Scrambling $h_t$ alters **41.2\%** of configuration decisions and causes nearly 9\% $Q$-value displacement. In contrast, permuting any non-network input (such as message size, active parameters, or GPU stats) changes at most 3.4\% of decisions.
\end{itemize}

Stripping away temporal network information changes agent behavior significantly. Together, Figures~\ref{eval:state_importance}(a) and~\ref{eval:state_importance}(b) demonstrate that configuration selector agent actively attends to dynamic network context, and this awareness directly governs configuration choices. This confirms that the network-adaptive design is the primary driver of performance gains under contention.

\section{Related Work}
\label{sec:related}

\textbf{Efficient Communication Algorithms.}
Topology-aware approaches such as Swing~\cite{swing} and MSCCLang~\cite{mscclang} construct collective communication algorithms tailored to specific physical topologies, while SCCL~\cite{sccl} synthesizes Pareto-optimal algorithms automatically. Compression-based methods reduce transferred data volumes using dynamic sparsity or quantization (e.g., HiPress~\cite{hipress}, ZeRO++~\cite{zero_pp}). The dynamic tuner presented in this work is orthogonal to these techniques: it optimizes runtime execution parameters for whichever collective algorithm is active, rather than altering the underlying algorithm or message payload itself.

\textbf{Overlapping Computation and Communication.} 
Pioneered by frameworks like Horovod~\cite{dp_horovod}, overlapping AllReduce collectives with backward pass computation reduces perceived communication overhead, with subsequent works extending this paradigm across hybrid parallelism strategies~\cite{comm_vs_comp_scaling_2, pccl}. The proposed tuner complements these scheduling techniques by directly reducing collective execution latency under network contention, allowing the performance benefits of both approaches to compose cleanly.

\textbf{Collective Communication Library Stacks and Tuning.}
Vendor and open-source stacks such as RCCL~\cite{rccl} and MSCCL~\cite{msccl} share foundational design elements with NCCL, while extensions like MSCCL++~\cite{msccl++}, NCCLX~\cite{ncclX}, and PCCL~\cite{pccl} introduce programmable kernels, optimized transport paths, and peer-to-peer execution abstractions.
For Mixture-of-Experts (MoE) workloads, NCCL EP~\cite{nccl_ep} provides specialized dispatch and combine primitives built directly on NCCL's Device API.
In the domain of configuration tuning, AutoCCL~\cite{autoccl} demonstrated the suboptimality of NCCL's default heuristics and motivated the NCCL Tuner Plugin API. Subsequent studies~\cite{nccl_tuning} have explored this configuration space, while AFNFA~\cite{afnfa} improves bandwidth prediction through offline profiling.
However, as highlighted in Section~\ref{sec:bkgrd_motiv}, existing autotuners suffer from measurement bias by optimizing for past or current observations rather than future network states, a limitation our dynamic tuner directly addresses through continuous runtime contention detection and proactive adaptation.

\textbf{Learning-Based Systems Optimization.}
Reinforcement learning (RL) has proven effective across diverse network and systems optimization tasks~\cite{deeprm, greennfv}.
While standard RL excels in static environments, capturing dynamic network behavior requires temporal context. Our dynamic tuner enhances standard RL by incorporating an LSTM layer to embed runtime network history directly into the RL agent's state representation. Combined with a novel multi-agent decomposition, this approach explicitly addresses the partially observable and temporally dependent nature of network contention, distinguishing our dynamic tuner from prior RL applications aimed at stateless systems.

\section{Conclusion}
\label{sec:conclusion}

Collective communication libraries have long operated under the assumption of a static, unperturbed network, fixing communication parameters at runtime based on idealized initial conditions. As distributed AI workloads scale across multi-tenant clusters, background traffic introduces stochastic, time-varying network contention that severely degrades performance. This work introduces a dynamic tuner that shifts collective communication from static, snapshot-based heuristics to continuous, stateful adaptation.

By decoupling contention detection (LSTM+SPRT agent) from configuration selection (D3QL agent), the proposed tuner provides a low-overhead mechanism to navigate the execution trade-offs of modern network fabrics like Slingshot-11. Our evaluation across six production AI workloads demonstrates that treating network state as a first-class control variable yields speedups of up to 65\% under network congestion, effectively recovering performance lost to static library heuristics.

Our results demonstrate that as LLM training and inference scale across shared, heterogeneous infrastructure, the communication stack must adapt dynamically alongside the workloads it serves. While this work focuses on contention-aware configuration tuning, its modular, multi-agent architecture provides a extensible framework to balance communication latency, energy efficiency, and in-network computing hardware such as NVIDIA SHARP or NVSwitch. We open-source our dynamic tuner to provide the community with a robust foundation for building autonomous, network-aware distributed communication systems.

\bibliographystyle{plain}
\bibliography{references}

%%%%%%%%%%%%%%%%%%%%%%%%%%%%%%%%%%%%%%%%%%%%%%%%%%%%%%%%%%%%%%%%%%%%%%%%%%%%%%%%
\end{document}